\documentclass[11pt]{article}
\usepackage{amsthm,amssymb, amsmath}
\usepackage[total={6.5in,8.8in}, top=1.1in, left=0.9in, includefoot]{geometry}

\usepackage{graphicx,epstopdf}
\usepackage[pdftex,bookmarks,breaklinks]{hyperref}

\renewcommand{\theenumi}{\alph{enumi}}

\renewcommand{\labelenumi}{\bf{\theenumi)}}

\newcommand{\Tset}{\mathbb{T}}
\newcommand{\Zset}{\mathbb{Z}}
\newcommand{\Nset}{\mathbb{N}}
\newcommand{\Rset}{\mathbb{R}}
\newcommand{\Sset}{\mathbb{S}}

\newcommand{\cC}{\mathcal{C}}
\newcommand{\cD}{\mathcal{D}}
\newcommand{\cE}{\mathcal{E}}
\newcommand{\cF}{\mathcal{F}}
\newcommand{\cG}{\mathcal{G}}

\newcommand{\cI}{\mathcal{I}}
\newcommand{\cL}{\mathcal{L}}
\newcommand{\cN}{\mathcal{N}}
\newcommand{\cP}{\mathcal{P}}
\newcommand{\cR}{\mathcal{R}}
\newcommand{\cS}{\mathcal{S}}
\newcommand{\cT}{\mathcal{T}}
\newcommand{\cU}{\mathcal{U}}

\theoremstyle{plain}
\newtheorem{teo}{Theorem}
\newtheorem{pro}[teo]{Proposition}
\newtheorem{lem}[teo]{Lemma}

\newtheorem{defi}{Definition}

\theoremstyle{remark}
\newtheorem*{remark}{Remark}

\newcommand{\codim}{\mathop{\rm codim}}
\newcommand{\close}[1]{\mathop{\rm cl}(#1)}
\newcommand{\interior}[1]{\mathop{\rm int}(#1)}
\newcommand{\volume}[1]{\mathrm{Vol}(#1)}
\newcommand{\area}[1]{\mathrm{area}(#1)}
\newcommand{\Diff}{\mathop{\rm Diff}}
\newcommand{\flux}{\mathop{\rm Flux}}
\newcommand{\volform}[1]{\Omega}
\newcommand{\Fix}[1]{\mathop{\rm Fix}(#1)}

\newcommand{\st}{{\rm s}}
\newcommand{\un}{{\rm u}}
\newcommand{\Ws}{W^\st}
\newcommand{\Wu}{W^\un}

\newcommand{\bn}{\mathbf{n}}

\newcommand{\pht}{\varphi_t}

\newcommand{\Sec}[1]{\S\ref{#1}}
\newcommand{\Thm}[1]{Thm.~\ref{#1}}
\newcommand{\Pro}[1]{Prop.~\ref{#1}}
\newcommand{\Lem}[1]{Lem.~\ref{#1}}

\newcommand{\Fig}[1]{Fig.~\ref{#1}}
\newcommand{\Def}[1]{Defn.~\ref{#1}}
\newcommand{\Eq}[1]{(\ref{#1})}    
\newcommand{\Hyp}[1]{\textbf{(H#1)}}

\title{Resonance Zones and Lobe Volumes for Volume-Preserving Maps}
\author{H.~E.~Lomel\'{\i} and J.~D.~Meiss \thanks
      {
        HL was supported in part by Asociaci\'{o}n Mexicana de Cultura.
        JDM was supported in part by NSF grant DMS-0707659 and by the
        Mathematical Sciences Research Institute in Berkeley.  Useful
        conversations with  Holger Dullin, Richard
        Montgomery and Rafael de la Llave are gratefully acknowledged.}
    \\
 \begin{tabular}{cc}
    Department of Mathematics         &         Department of Applied Mathematics\\
        Instituto Tecnol\'{o}gico Aut\'{o}nomo de M\'{e}xico  &University of Colorado \\
    Mexico, DF 01000                    &       Boulder, CO 80309-0526 \\
    lomeli@itam.mx                      &       James.Meiss@colorado.edu\\
\end{tabular}
}

\begin{document}
\maketitle

\begin{abstract}
We study exact, volume-preserving diffeomorphisms that have
heteroclinic connections between a pair of normally hyperbolic invariant
manifolds. We develop a general theory of lobes, showing that
the lobe volume is given by an integral of a generating form over the primary intersection,
a subset of the heteroclinic orbits. Our definition reproduces the classical action formula in the planar, twist map case. For perturbations from a heteroclinic connection, the lobe volume is shown to reduce, to lowest order, to a suitable integral of a Melnikov function.


\end{abstract}

\section{Introduction}

The computation of the volume of incoming and exit sets for a ``nearly invariant''
region is the first step in the development of a dynamical theory of
transport. For area-preserving maps, it is common for these regions to
be bounded by segments of the stable and unstable manifolds hyperbolic invariant
sets, typically periodic orbits or cantori (Aubry-Mather sets) \cite{MMP84,
RomKedar88}. In this case the resulting set is a ``resonance zone'' \cite{MMP87,
Easton91}, and its exit and entrance sets are ``lobes'' of the ``turnstile.'' When the splitting between the manifolds is small, the resonance zone is nearly
invariant.

The volume of a lobe is a not only a measure of the separation of the stable and
unstable manifolds, it is also the \emph{flux} of trajectories escaping from the
resonance zone. This flux provides an estimate for the escape time from the
resonance. It is known that the average exit time from a region is equal to the ratio
of accessible volume of the region (the fraction of the volume that can be reached
by orbits from the outside) to the flux \cite{MacKay94, Meiss97}. Though the
accessible volume is difficult to compute it is certainly bounded by the total
volume of the region, so the flux provides an upper bound on the average exit
time. Moreover, if the flux goes to zero, but the accessible volume does not, then
the average exit time must go to infinity.

In this paper, we discuss the construction of lobes for resonance zones of
volume-preserving maps. Our goal is to generalize the results of \cite{MMP84} that
provide formulas for the lobe areas for the two-dimensional maps and of
\cite{MacKay94} for the case of three-dimensional, incompressible vector fields.
Indeed, one of the three open problems posed by MacKay at the end of
\cite{MacKay94} is to generalize his flow results to the case of maps.

The theory of \cite{MacKay94} applies to exact-symplectic maps. 
Recall that a map is symplectic when there is a
closed two form, $\omega$, (for example $\omega = dq \wedge dp$) that is preserved
by $f$: $f^*\omega = \omega$.\footnote 
{We recall the notation for the pullback, $f^*$, and similar concepts in the Appendix.}
A map is exact-symplectic when $\omega$ is exact, $\omega = -d \nu$ (for example
$\nu = p dq$) and there exists a function $S$ defined by
\begin{equation}\label{eq:exactSymplectic}
    f^* \nu - \nu = dS \;.
\end{equation}
An often studied case is that of twist maps, which have Lagrangian generating
functions---the discrete analogues of the Lagrangians for differential equations
\cite{Meiss92}.

Remarkably, the generating function \Eq{eq:exactSymplectic} provides a way to
compute the area of a lobe (or of the resonance zone itself) in terms of the
action, the formal sum of $S$ along an orbit \cite{MMP84, MMP87}. The result is
that a two-dimensional integral over a lobe is reduced to a zero-dimensional
integral, the difference between the actions of the orbits homoclinic to the
hyperbolic invariant set. Thus to compute the lobe area one only needs to find
these homoclinic orbits and carry out the sum; this is considerably easier than constructing
the entire lobe boundary and computing the two-dimensional integral. Moreover,
since the action is stationary on the orbits, its computation is second-order
accurate.

For the case of incompressible vector fields, a similar result also holds. On the manifold $M =\Rset^3$, when $\nabla \cdot u = 0$ there exists a vector potential $A$ so that $u = \nabla \times A$. Here $A$ is more properly thought of as a
one-form $\beta = A \cdot dl$ and it gives rise to a variational principle for
orbits of the flow. Moreover, MacKay has shown that the one-form $\beta$ can be
used to compute the flux through two-dimensional surfaces  \cite{MacKay94}. More
generally, when $\volform{M}$ is an exact volume form and $u$ is an {\em exact}
incompressible vector field, then the curl relationship generalizes to 
the statement that $i_u \volform{M} = d\beta$ is exact and $\beta$ provides 
the generating form (see \Sec{sec:Hamiltonian} and \cite{Lomeli08b}) as well 
as a variational principle\cite{Gaeta03}.

In this paper we study exact volume-preserving maps. For such maps there is also
an analogue of \Eq{eq:exactSymplectic} in which the generator $S$ becomes a
differential form $\lambda$, see \Sec{sec:exact}. Such forms were implicitly
described by \cite{Carroll04} in the context of constructing implicit generating
functions for volume-preserving maps \cite{Lomeli08b}. We will primarily study the
three-dimensional case where the generator $\lambda$ is a one-form.

We will construct a resonance zone based on a pair of normally hyperbolic
invariant manifolds, say $A$ and $B$. For the three-dimensional case, 
the natural objects are periodic orbits and invariant circles. 
The boundary of the resonance will consist of pieces
of the codimension-one stable $\Ws(A)$ and unstable $\Wu(B)$ manifolds of these
invariant sets. The exit and incoming sets are obtained by iterating the
resonance, see \Sec{sec:transport}.

Unlike the two-dimensional case, the boundary of a resonance does not always
consist solely of pieces of stable and unstable manifolds. Indeed for this to
happen, the manifolds would have to intersect on a ``proper boundary'', see
\Sec{sec:fundamental}. However, we have observed that the set of heteroclinic
intersections commonly contains components that are themselves bi-asymptotic to
the invariant sets; these preclude the existence of heteroclinic proper boundaries
\cite{Lomeli00a}. A similar phenomena occurs for three-dimensional
volume-preserving flows: when the two-dimensional manifolds of saddle equilibria
intersect, they typically do so only along a few ``primary" heteroclinic orbits
\cite{Broer81, Holmes84, MacKay94}. To remedy this, a ``cap'' must be used to
complete the resonance boundary. Even though the cap is---to a large
extent---arbitrary, this choice does not change the volume of the exit and
incoming lobes of the resonance.

Our main result is \Thm{thm:main} in \Sec{sec:lobeVolume}, which states that the
volume of the exit and incoming lobes is given by
\[
    \volume{\cE} = \volume{\cI} 
              =\sum_{k\in\mathbb{Z} }\int_\eta (f^k)^*\lambda
              =\int_{\cP(A,B)}\lambda \;.
\]
Here the set $\cP(A,B)=\bigcup_{k\in\Zset}f^k(\eta)$ consists of the primary 
heteroclinic intersections of the
manifolds $\Ws(A)$ and $\Wu(B)$, and $\eta$ is the restriction of
$\cP(A,B)$ to a fundamental domain. Thus to compute the volume of
$n$-dimensional lobes one needs only to do $(n-2)$-dimensional integrals along
submanifolds of heteroclinic intersections.

For the case of nearly-integrable systems, a widely used technique for detecting
such heteroclinic intersections is the Poincar\'{e}-Melnikov method
\cite{Melnikov63,HolmesM82,Lomeli08a}. Indeed, this method detects---at first
order---precisely the primary intersections.

The classical Melnikov function computes the rate at which the distance between
the manifolds changes with a perturbation, say $\delta$. For two-dimensional maps,
an integral of the Melnikov function between a pair of zeros gives the rate of
change of the flux with $\delta$ \cite{MacKay88}; a similar result also holds for
incompressible vector fields \cite{MacKay94}. We will show in \Sec{sec:melnikov}
that this same results holds for volume-preserving maps.

We conclude by presenting several examples and applications in \Sec{sec:examples}.

\section{Exact Volume-preserving maps} \label{sec:exact}

\subsection{Definition}

Let $M$ be an $n$-dimensional manifold.
A volume form $\volform{M}$ is a nondegenerate $n$-form on $M$.
A map $f: M \to M$ on an $n$-dimensional manifold preserves the volume form $\volform{M}$
if
\[
    f^*\volform{M} = \volform{M} \;.
\]
For example if $M = \Rset^n$, and  $\volform{M} = dx_1 \wedge dx_2 \wedge \ldots \wedge
dx_n$, then $f$ is volume preserving when its Jacobian has unit determinant everywhere,
$\det(Df) = 1$. Suppose now that $\volform{M}$ is exact, i.e., there exists an
$(n-1)$-form $\alpha$ such that $\volform{M}= d\alpha$. By analogy with the symplectic
case \Eq{eq:exactSymplectic}, we can also define exact volume-preserving maps.

\begin{defi}[Exact Volume Preserving]\label{def:exact}
A diffeomorphism $f:M\to M$ is \emph{exact}-volume preserving if there exists an
$(n-1)$-form $\alpha$ such that $d\alpha = \volform{M}$ and a \emph{generating}
$(n-2)$-form $\lambda$ such that
\begin{equation}\label{eq:exact}
 f^*\alpha-\alpha= d\lambda.
\end{equation}
\end{defi}

It is clear that if $f$ is exact-volume preserving, then $f^{-1}$ is also.
Moreover, if $f = g_1 \circ g_2$ is the composition of exact volume-preserving
maps with generating forms $\lambda_1$ and $\lambda_2$, respectively, then since
$(g_1 \circ g_2)^* = g_2^* \circ g_1^*$,
\[
    f^*\alpha - \alpha = g_2^* (g_1^*\alpha -\alpha) + g_2^*\alpha -\alpha
       = d(g_2^*\lambda_1 +\lambda_2) \;.
\]
Thus $f$ is exact-volume preserving with $
    \lambda = g_2^*\lambda_1 + \lambda_2 \;.
$ Consequently, the set of exact volume-preserving maps is a subgroup of the group
of volume-preserving diffeomorphisms, we denote it $\Diff_\alpha(M)$. We discuss
some of the consequences of exactness in \cite{Lomeli08b}.

As an application that we will need later, consider the composition of an exact
volume-preserving map with itself. In this case we can use a telescoping sum to
conclude
\[
    (f^n)^* \alpha - \alpha =  \sum_{j=0}^{n-1}\left((f^{j+1})^*\alpha -(f^j)^*\alpha \right)
    = \sum_{j=0}^{n-1} (f^{j})^* \left(f^*\alpha -\alpha \right)
    = \sum_{j=0}^{n-1} d((f^j)^*\lambda) \;.
\]
Thus $f^n$ is exact-volume preserving with the form
\begin{equation}\label{eq:composition}
    \lambda_n = \sum_{j=0}^{n-1}  (f^j)^*\lambda \;.
\end{equation}

\subsection{Geometrical Implications}
Exact volume-preserving maps arise naturally in the context of perturbations from
integrable maps. For example suppose that $f$ is a map on the phase space $M =
\Tset^d \times \Rset$ with volume form $\volform{M} = dz \wedge d\theta_1 \wedge \ldots
\wedge d\theta_d$. This form is exact with $\alpha = z \wedge d\theta_1 \wedge
\ldots \wedge d\theta_d$. One integrable map on $M$ is
\begin{equation}\label{eq:integrable}
    f(\theta,z) = (\theta + \rho(z), z) \;,
\end{equation}
with the rotation vector $\rho : \Rset \to \Tset^d$. This map is also exact volume preserving: if we
define the vector field $ W(\theta,z) =(0, \int z \rho' dz)$, then $f^*\alpha-\alpha=d\lambda$ where
\[
    \lambda =  i_W d\theta_1 \wedge \ldots \wedge d\theta_d \;.
\]
The invariant tori of this map are ``rotational'' tori, and as we argue next, any
volume preserving map on $M$ with a rotational invariant torus must be exact.

A \emph{rotational} torus on $\Tset^d \times \Rset$ is a $d$-dimensional torus
that is homotopic to the zero section $\{ (\theta,0): \theta \in \Tset^d \}$. The
\emph{net flux} crossing a rotational torus $\cT$ is the difference between the
signed volume below $f(\cT)$ and that below $\cT$:
\begin{equation}\label{eq:netFlux}
    \flux(\cT) = \int_{\cT} f^*\alpha - \alpha \;.
\end{equation}
Note that if $f$ is volume preserving, and $\cT$ and $\hat\cT$ are any two
rotational tori then the volume contained between them, $\Delta V = \int_{\hat\cT}
\alpha - \int_{\cT} \alpha$, is invariant. This implies that $\flux(\cT)$ is independent
of the choice of torus.

Thus, if  $\flux(\cT) = 0$ then $f(\cT) \cap \cT \neq \emptyset$ for any rotational
torus; this \emph{intersection property} is used in the generalization of KAM
theory for exact volume-preserving maps \cite{Xia92}. Conversely, if $f$ has an
invariant rotational torus then $\flux(\cT)= 0$ by definition. Consequently if $f$ has a
rotational invariant torus then it is necessarily exact.

Exactness can be used to simplify  the computation of the volume of certain
regions. Suppose that $M$ is any orientable manifold, and $\cR$ is a region whose
boundary can be decomposed into pieces  that are related by iteration, $\cS$ and
$f(\cS)$, and such that $\cS \cap f(\cS) =\cC$ is an invariant codimension-two
submanifold. We assign orientations to $\cR$ and $\cS$, and by iteration to
$f(\cS)$; thus
\[
    \partial \cR =  f(\cS) - \cS \;,
\]
and $\partial \cS = \cC$. Then, when $f$ is exact, \Eq{eq:exact} immediately gives
\[
    \volume{\cR} = \int_\cR \volform{M} = \int_\cS f^*\alpha -\alpha = \int_{\cC} \lambda \;.
\]
Thus we can compute the volume of $\cR$ simply by integrating over $\cC$. This
formula is closely related to those that we will use to compute the volume of exit
and incoming sets in \Sec{sec:lobeVolume}.

\subsection{Examples}
Since every closed form on $M=\Rset^n$ is exact, the volume form $\volform{M} = dx_1
\wedge \ldots \wedge dx_n$ is exact with, for example, $\alpha = x_1 dx_2 \wedge
\ldots \wedge dx_n$. If $f$ is volume preserving, then $f^*\alpha -\alpha$ is  a
closed $(n-1)$-form, but since every closed form on $\Rset^n$ is exact, $f$ is exact.

However, exactness is not automatic on more general manifolds. For example, a
perturbation of \Eq{eq:integrable} is the one-action map $f$ on $M=\Tset^d,
\times \Rset$
\begin{equation}\label{eq:oneAction}
    f(\theta, z) = \left(\theta + \rho(z + F(\theta)) , z + F(\theta) \right) \;,
\end{equation}
for a ``force'' $F: \Tset^d \to \Rset$ is always volume-preserving, but is only
exact when the form $F d\theta_1 \wedge \ldots \wedge d\theta_d$ is exact
\cite{Lomeli08b}, or equivalently when
\[
    \int_{\Tset^d} F d\theta_1 \wedge \ldots \wedge d\theta_d = 0 \;.
\]
For example, if $d=2$, $\rho(z) = (z, z^2)$ and $F = a \cos \theta_1 + b \cos
\theta_2 + c \cos(\theta_1 + \theta_2)$ then \Eq{eq:oneAction} is generated by
\[
    \lambda = \left(\frac12 Z^2 + a \sin \theta_1 \right) d \theta_2 -
              \left (\frac23 Z^3 + b \sin \theta_2 +
                 c \sin(\theta_1 + \theta_2)\right) d\theta_1 \;,
\]
where $Z = z+ F(\theta)$.

In \cite{Lomeli08b} we showed that any exact-symplectic map
\Eq{eq:exactSymplectic} is also exact-volume preserving. For example, reinterpreting
\Eq{eq:oneAction} as a map on $(\theta,z) \in \Tset^2 \times \Rset^2$, then $f$
becomes a ``generalized standard'' or  ``Froeschl\'e'' map \cite{Froeschle72}. It
is symplectic with the two-form $\omega = \sum_i d\theta_i \wedge dz_i$ provided
that $\rho(z) = \nabla K(z)$ for ``kinetic energy'' $K$ and $F = -\nabla
V(\theta)$ for ``potential energy'' $V$.  Letting $\nu = z \cdot d\theta$, then
$f$ is also exact-symplectic and has the generator
\[
    S(\theta,z) = Z \cdot \nabla K(Z) - K(Z) -V(\theta) \;,
\]
which is equivalent to the implicit, Lagrangian generating function
\cite{Meiss92}. For example when $\rho(z) = z$, then $K(z) = \frac12 |z|^2$ and
the generator becomes 
\[S(\theta,z) =\frac12 |Z|^2 -V(\theta).\]  Now setting
$\alpha = \frac12 \nu \wedge \omega$, then $d \alpha = \volform{M} = d\theta_1 \wedge d
\theta_2 \wedge dz_1 \wedge dz_2$ and  we find that the generator \Eq{eq:exact}
for the generalized standard map is the two-form
\[
    \lambda = \frac12 S(\theta,z) \omega \;.
\]

\section{Normally hyperbolic invariant manifolds}

Resonance zones are most naturally associated with normally hyperbolic invariant manifolds; in the simplest case, with a hyperbolic periodic orbit. In addition to periodic orbits, we will also consider normally hyperbolic invariant circles, but many of our results apply more generally. Here we recall the definition of normal hyperbolicity, and prove a lemma about convergence of forms on the stable manifolds of such sets.

Suppose and $f : M\to M$ is a diffeomorphism on an $n$-dimensional smooth
manifold $M$ and $A$ is a compact invariant set of $f$. We recall one standard definition
of normal hyperbolicity.

\begin{defi}[Normal Hyperbolicity \cite{DDS08}] \label{defi:normhyp} A compact invariant set $A$ is \emph{$r$-normally hyperbolic} for $r \in \Nset$, if there exists an invariant splitting of the tangent bundle $T_AM= E^\st\oplus E^\un\oplus T A$, a Riemannian structure, and
positive constants $C,\lambda$ and $\mu$, such that for all $a \in A$,
\begin{enumerate}
 \item $0 < \lambda < \mu^{-r} < 1$;
 \item $\|Df^t(a)v\|\le C \lambda^n\|v\|$, for all $v\in E^\st_a$ and $t\in\Nset$;
 \item $\|Df^{-t}(a)v\|\le C \lambda^n\|v\|$, for all $v\in E^\un_a$ and $t\in\Nset$;
 \item $\|Df^t(a)v\|\le C \mu^{|n|}\|v\|$, for all $v\in T_aA$ and $t\in\Zset$.
\end{enumerate}
\end{defi}

The stable manifold theorem applies to $r$-normally hyperbolic sets: 
there exist $C^r$ immersed submanifolds $\Ws(A,f) = \Ws(A)$, tangent to $E^\st\oplus TA$ at $A$, and $\Wu(A,f) = \Wu(A)$, tangent to $E^\un\oplus TA$ at $A$ \cite{HirschPS77}. 

We next prove a lemma that we will need in the following sections to show that some of the sums converge. This lemma applies to differential $\ell$-forms on a compact part of a stable or unstable manifold of a normally hyperbolic invariant set.
 
Let $\Lambda^\ell(M)$ denote the linear space of $\ell$-forms on $M$. For any compact set $P \subset M$, there is a natural norm on $\Lambda^\ell(P)$ given by:
\[
	\|\omega\|_P=\sup\left\{|\omega_p(v_1,v_2,\ldots,v_\ell)|: 
	      p\in P,\; v_i \in T_p P,\; \|v_i\|=1,\; i = 1, \dots, \ell \right\}.
\]
The main point is that for any $p \in P$ and any set of vectors $v_i \in T_p P$,
\[
|\omega_x(v_1,v_2,\ldots,v_\ell)|\le \|\omega\|_P \,\|v_1\|\,\|v_2\|\cdots\|v_\ell\| \;.
\]

\begin{defi}[Regular Form]
A differential $\ell$-form $\omega$ defined on $\Ws(A)$ is \emph{regular} if
\[
	\lim_{k\to\infty}(f^*)^k \omega=0 \;.
\]
Similarly if $\omega$ is defined on $\Wu(A)$ then it is \emph{regular} if
\[
	\lim_{k\to -\infty}(f^*)^k \omega=0 \;.
\]

\end{defi}

In \Sec{sec:lobeVolume} and \Sec{sec:melnikov}, we will use the following simple consequences of this definition.

\begin{lem}\label{lem:aux2}
If $\omega$ is regular on $\Ws(A)$, then
\begin{enumerate}
\item if $P\subset\Ws(A)$ is a compact submanifold of dimension $\ell$ then
\[
	\lim_{t\to\infty}\int_P  (f^t)^* \omega=0 \;;
\]
\item If $Y$ is a vector field on $\Ws(A)$ 
and $Q\subset\Ws(A)$ is submanifold of dimension $\ell-1$  then
\[
	\lim_{t\to\infty}\int_Q i_Y (f^t)^* \omega=0 \;.
\]
\end{enumerate}
\end{lem}

An $\ell$-form is automatically regular if $\ell$ is large enough so that every independent set of $\ell$ vectors $v_i \in T\Ws(A)$ is guaranteed to contain sufficiently many vectors in the stable, as opposed to center, directions.

\begin{lem}\label{lem:aux1}
If $A$ is an $r$-normally hyperbolic invariant set with $\dim(A) = n_A$, then every differential $\ell$-form with $\ell > 0$ and
\begin{equation}\label{eq:ellRegular} 
	\ell \ge n_A\left(1+\frac{1}{r}\right)
\end{equation}
is regular on any compact subset $P \subset \Ws(A)$.
\end{lem}

\proof
By hypothesis there is a splitting \(T_AM= E^\st\oplus E^\un\oplus T A \) and
positive constants $C,\lambda$ and $\mu$ as in \Def{defi:normhyp}. 
The stable manifold theorem implies that there exists a neighborhood $\cN$ of $A$ in $\Ws(A)$ and coordinates $\phi=(a,s):\cN\to A\times \Rset^{n_s}$, where $n_s=\dim(E^\st)$, such that
$s(\xi)=0$ if and only if $\xi\in A$, and such that $\tilde{f}=\phi\circ f\circ \phi^{-1}$ takes the form
\[
	\tilde{f}(a,s)=(g(a), L(a)s+ r(a,s))
 \]
with $r(a,0)=0$, $\partial_2 r(a,0)=0$  and $L(a)$ is a matrix.
In addition, there is an induced Riemannian 
structure on $\cN$ such that $D\tilde{f}$ satisfies the conditions of normal hyperbolicity 
on the zero section, for some constants $\tilde C$, $\tilde{\lambda}$ and $\tilde{\mu}$, such that $\lambda \le \tilde{\lambda}<1$, $1<\tilde{\mu} \le \mu$, and $\tilde{\mu}^r\tilde{\lambda}<1$.

Thus if $P$ is a compact submanifold of $\Ws(A)$, the normal hyperbolicity conditions imply that for all $t\in\Nset$ there exists $m_1,m_2 \in \Nset$ such that
\[
 	\|(f^t)^*\omega\|_P \le 
	  \tilde{C}\left(\tilde{\mu}^{m_1}\tilde{\lambda}^{m_2}\right)^t\|\omega\|_P \;.
\]
where $m_1+m_2 = \ell$ and $m_1 \le n_A$.
The $r^{th}$ power of the contraction factor in this equation satisfies
\[
	 \left(\tilde{\mu}^{m_1}\tilde{\lambda}^{m_2} \right)^r
	= \left(\tilde{\mu}^{r}\tilde{\lambda} \right)^{m_1}\tilde{\lambda}^{(m_2r-m_1)} \;,
\]
and \Eq{eq:ellRegular} implies that 
\[
	m_2r-m_1=(\ell-m_1)r-m_1\geq (\ell-n_A)r -n_A\geq 0 \;.
\]
Thus both integers $m_1$ and $m_2r-m_1$ are nonnegative. If at least one of these integers is positive, then $ \tilde{\mu}^{m_1}\tilde{\lambda}^{m_2} <1$. If $m_1 > 0$ then all is well. Alternatively, if $m_1 = 0$, then $m_2r -m_1 = \ell r$, which is positive since $\ell > 0$ by hypothesis. Consequently, in either case, $ \|(f^n)^*\omega\|_P\to 0$ as required.

\qed

\begin{remark}
It is important to notice that we have to include the manifold $A$ in $\Ws(A)$, otherwise we
could have invariant forms on $\Ws(A)$ that don't satisfy the conclusion of
\Lem{lem:aux1}. For instance, in \cite{Lomeli03}, an invariant $n-1$ form is constructed on $\Ws(A) \setminus A$ when $A$ is a normally hyperbolic invariant circle. We also want to point out that similar computations appear in \cite{DDS08} and \cite{Cabre05}.
\end{remark}

If $n_A = 0$, then \Lem{lem:aux1} implies that all $\ell$-forms with $\ell \ge 1$ are regular, but if $n_A = 1$, then $\ell \ge 2$. We will often apply the lemma to the case $\ell = n-1$, in which case an $\ell$-form always is regular on a one-normally hyperbolic invariant manifold when $n_A \le (n-1)/2$.

\section{Measuring Transport}\label{sec:transport}

In dynamical systems, transport is the study of the motion of collections of
trajectories from one region of phase space to another. A natural choice for a
region is a subset $\cR\subset M$  that is \emph{almost invariant} in the sense that
it consists of a set of points whose orbits belong to $\cR$ for a long time.
One way to construct such a nearly invariant region is to form its boundary
$\partial\cR$ (as much as possible) from the invariant manifolds of a pair of
normally hyperbolic invariant manifolds; these boundaries are called \emph{partial barriers}.\footnote
{
	Another is to use an isolating block \cite{Easton91}
}

\begin{figure}[th]
\begin{center}
\includegraphics[height=3in]{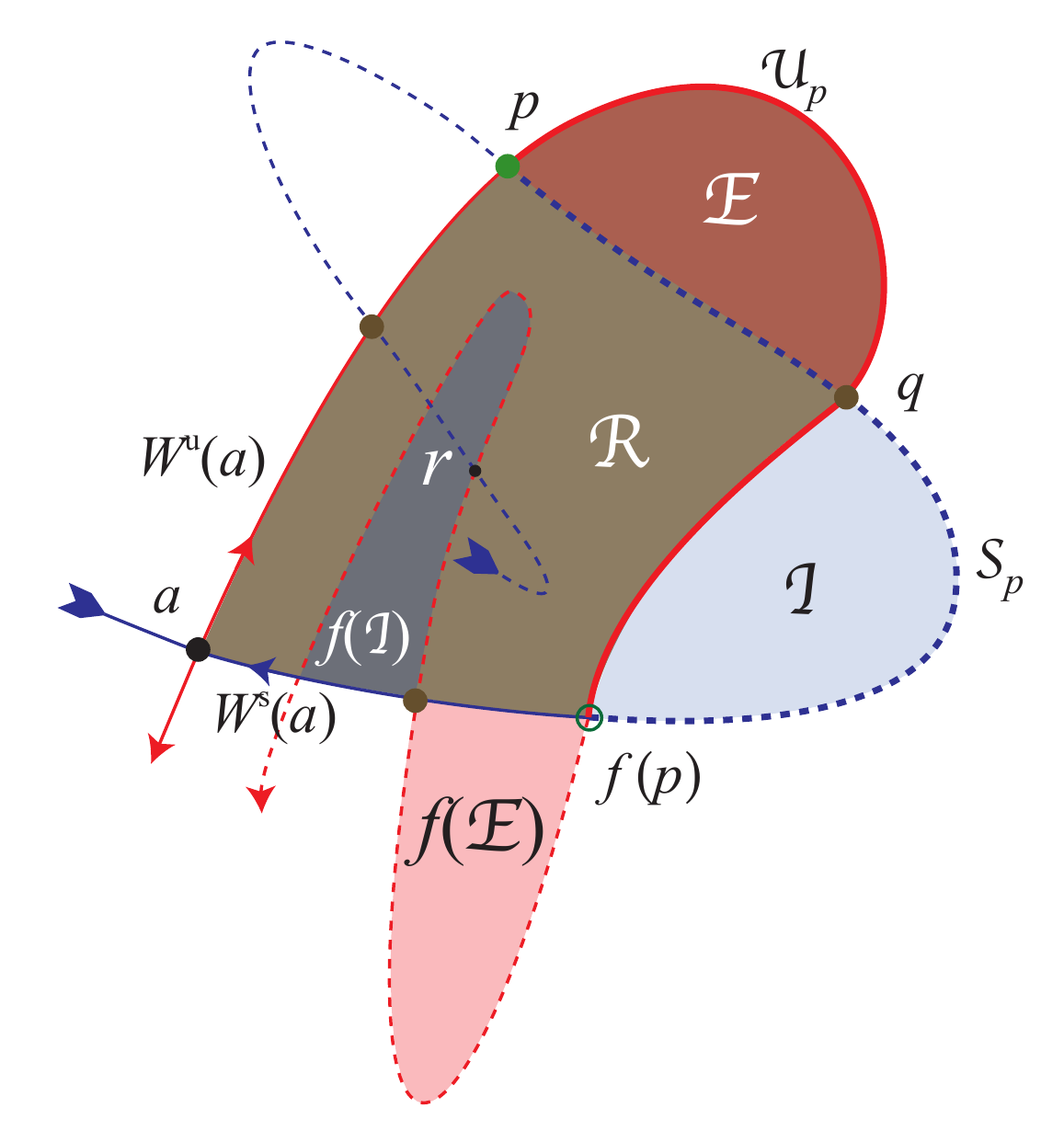}
\caption{Standard construction of a two-dimensional resonance zone $\cR$ for a
saddle point $a$, based on a primary homoclinic point $f(p) \in \Ws(a) \cap
\Wu(a)$. The  boundary of the resonance zone $\cR$ is 
$\Wu_{f(p)}(a) \cup \Ws_{f(p)}(a)$ and
$\cR$ has exit set $\cE$ and incoming set $\cI$.} \label{fig:2DResonanceStd}
\end{center}
\end{figure}

For the two-dimensional case this leads to the construction of \emph{resonance
zones} bounded by the invariant manifolds of a periodic orbit or cantorus
\cite{MMP84, Easton91}. For a saddle fixed point, $a$, it is natural to base the
selection of the exit and incoming sets on primary homoclinic points $p,q \in
\Ws(a) \cap \Wu(a)$, see \Fig{fig:2DResonanceStd}. In this case the boundary of
the resonance is a single partial barrier formed from segments of the stable and
unstable manifolds from $a$ to $f(p)$. In other cases, like maps on the cylinder,
the resonance zones associated with rotational periodic orbits have upper and
lower boundaries, each formed from different branches of stable and unstable
manifolds \cite{MMP87}.

In this paper we will generalize this construction to the volume-preserving case.
To do this, we will not be able to assume that the lobes are bounded entirely by
subsets of the invariant manifolds. 

In order to restrict the topological possibilities to
something manageable, we will usually assume that the phase space $M$ is at most
three-dimensional and consider only the simplest hyperbolic invariant sets:
fixed points and invariant circles.

\subsection{Incoming and Exit Sets}\label{sec:exitSet}

Any region $\cR \subset M$ has \emph{incoming} and \emph{exit} sets. The exit set
$\cE$ is the set of points in $\cR$ that leave in one step: they have forward exit
time one. Similarly, the incoming set $\cI$ consists of all points not in $\cR$
whose images land in $\cR$.\footnote
{
    In \cite{Meiss97}, the entry set was defined to be the image of the incoming set
    and is a subset of $\cR$. Here it seems more convenient to use the preimage of
    the entry set.
}
\begin{defi}[Incoming and exit sets] The \emph{exit} and \emph{incoming} sets for a region $\cR$ are
\begin{equation}\label{eq:sets}
\begin{split}
  \cE &= \cR\setminus f^{-1}(\cR) \;, \\
  \cI &= f^{-1}(\cR) \setminus \cR \;,
\end{split}\end{equation}
respectively.
\end{defi}

Suppose now that $f$ preserves a measure $\mu$, and that $\mu(\cR) < \infty$.
Since $\mu(A \setminus B) = \mu(A) - \mu(A \cap B)$, then
\begin{align*}
    \mu(\cE) &= \mu(\cR) -\mu(\cR\cap f^{-1}(\cR)) \;, \\
    \mu(\cI) &= \mu(f^{-1}(\cR)) -\mu(f^{-1}(\cR)\cap \cR) \;.
 \end{align*}
My measure preservation, $\mu(\cR )=\mu(f^{-1}(\cR))$, so that the volume of the
exit and incoming sets are equal: $\mu(\cI) = \mu(\cE)$.

The incoming and exit sets for a region are often called ``lobes''
\cite{RomKedar88}, and their union is a ``turnstile'' \cite{MMP84}.  The volume of
a lobe---either exit or incoming set---is a simple measure of how fast the points
move from $\cR$ to its complement. In other words, the lobe volume is a measure of
the degree to which a set is  invariant. Roughly speaking, if one selects a point
at random in $\cR$, then it has probability $p = \mu(\cE)/\mu(\cR)$ of landing in
$\cE$ at each iteration, and so its expected escape time is $1/p$. This can be
made precise \cite{Meiss97}: the average time for a trajectory that begins in
$\cI$ to escape is
\[
    \left.\langle t_{exit} \rangle\right|_{\cI} = \frac{ \mu(\cR_{acc})} {\mu(\cE)} \;,
\]
where $\cR_{acc}$ is the \emph{accessible} portion of $\cR$: the part that can be
reached by trajectories that begin outside $\cR$. To obtain more details of the
distribution of exit times requires an understanding of the decomposition of $\cI$
into regions of fixed exit time \cite{RomKedar88, Meiss97}.

\subsection{Fundamental Domains}\label{sec:fundamental}

To construct the boundaries of a resonance zone, we will use the concepts of
\emph{fundamental domain} and \emph{primary intersection} of the stable and
unstable manifolds of a hyperbolic invariant set \cite{Lomeli00a}.

In order to form a resonance zone based upon a pair of normally 
hyperbolic invariant manifolds $A$ and $B$, they must have 
codimension-one invariant stable and unstable manifolds. To be concrete, we
will suppose that
\begin{equation}\label{eq:dimension}
    \codim(\Ws(A)) = \codim(\Wu(B)) = 1 \;.
\end{equation}

If $\sigma \subset \Ws(A)$ is the boundary of a subset of $\Ws(A)$ that
contains $A$, then we denote the unstable manifold \emph{starting at} $\sigma$ by
$\Ws_\sigma(A)$: it is the \emph{closed} subset of the local stable manifold of
$A$ bounded by $\sigma$:
\[
    \sigma = \partial W^s_\sigma(A) \;.
\]
Similarly, $\Wu_\gamma(B)$ is the unstable manifold \emph{up to} $\gamma$ when
$\gamma = \partial \Wu_\gamma(B)$; however, in this case it is convenient to
assume that this submanifold is \emph{open}. Though this definition is not not
symmetric, the asymmetry is useful to simplify some proofs.

For our purposes, it will be important that the boundaries of these local
manifolds are chosen to be \emph{proper}. Recall that a neighborhood $N$ of an
invariant set $A$ is \emph{isolating} or a \emph{trapping region} if
$A\subset \interior{N}$ and
\[
    f(\close{N}) \subset \interior{N}\;.
\]

\begin{defi}[Proper Boundary]\label{def:guasa}
Suppose $A$ and $B$ are compact, normally hyperbolic invariant manifolds. A set
$\sigma \subset \Ws(A)$ is a \emph{proper boundary} if $\Ws_\sigma(A)$ is an
isolating neighborhood of $A$ in $\Ws(A)$. Similarly, $\gamma \subset \Wu(B)$ is
\emph{proper} if $\Wu_\gamma(B)$ is an isolating neighborhood of $B$ for
$f^{-1}$.
\end{defi}

\noindent It is not hard to see that the stable manifold theorem, e.g.
\cite{HPS77}, implies that proper boundaries always exist. These local manifolds
behave naturally under iteration:
\[
    f(\Ws_{\sigma}(A)) = \Ws_{f(\sigma)}(A) \;, \mbox{  and  }
    f(\Wu_{\gamma}(B)) = \Wu_{f(\gamma)}(B) \;.
\]
Given a proper boundary, an invariant manifold can be partitioned into
nonoverlapping {\em fundamental} domains that are related by iteration.

\begin{defi}[Fundamental Domain]
The set $\cF^\st(A)$ of fundamental domains of $\Ws(A)$ is the collection of sets
of the form
\[
    \cS_\sigma(A) \equiv W_{\sigma}^\st(A)\setminus W_{f(\sigma)}^\st(A) \;.
\]
where $\sigma$ is any proper boundary. Similarly, the set $\cF^\un(B)$ of
fundamental domains in $\Wu(B)$ is the collection of sets of the form
\[
    \cU_\gamma(B) \equiv W_{f(\gamma)}^\un(B)\setminus W_{\gamma}^\un(B) \;,
\]
where $\gamma$ is any proper boundary.
\end{defi}

\begin{remark}
The closure assumptions imply that $\sigma \subset \cS_\sigma$ and
$\gamma \subset \cU_\gamma$, but their images are not. In addition,
one has $\partial \cS_\sigma=\sigma \cup f(\sigma)$ and
$\partial \cU_\gamma=\gamma \cup f(\gamma)$.
\end{remark} 

As an example, consider the two-dimensional map sketched in \Fig{fig:2DResonanceStd}.
In this case, the resonance
zone $\cR$ is bounded by $\Ws_{f(p)}(a) \cup \Wu_{f(p)}(a)$ and the
fundamental domains $\cS_p(a)$ and $\cU_p(a)$ form the boundary of the region $\cE
\cup \cI$, the turnstile for $\cR$. For a three-dimensional map, a fundamental
domain is an annulus bounded by a proper boundary $\gamma$ and its image
$f(\gamma)$.

A consequence of the definition is that the image of a fundamental domain is also
a fundamental domain and that $f^{k}(\cS_\sigma) = \cS_{f^{k}(\sigma)}$ and
$f^{k}(\cU_\gamma) = \cU_{f^{k}(\gamma)}$ for any $k \in \Zset$. Moreover, stable
and unstable manifolds can be decomposed as the disjoint union of fundamental
domains:
\begin{equation}\label{eq:decomp}
\begin{split}
     \Ws_\sigma(A)\setminus A &= \bigcup_{t \ge 0} \cS_{f^t(\sigma)}(A) \;,\\
     \Wu_\gamma(B)\setminus B &= \bigcup_{t < 0} \cU_{f^t(\gamma)}(B) \;.
\end{split}\end{equation}
Consequently, the topology of the intersections of stable and unstable manifolds
can be studied by restricting to appropriate fundamental domains \cite{Lomeli00a,
Lomeli03}.  We will use the fundamental domains to construct the incoming and exit
sets for a resonance zone.

\subsection{Primary Intersections}\label{sec:primary}

As above, we continue to assume that $f$ has a pair of normally hyperbolic
invariant sets $A$ and $B$. In addition, the stable and unstable manifolds
$\Ws(A)$ and $\Wu(B) $ are  codimension-one, orientable manifolds as in \Eq{eq:dimension}. 
In addition, we now
assume that there is a heteroclinic intersection
\begin{equation}\label{eq:intersect}
    \Ws(A) \cap \Wu(B) \neq \emptyset \;.
\end{equation}
The set of heteroclinic intersections is typically staggeringly complex;  in this
section we pick out the first or \emph{primary} intersection to use in the
construction of a partial barrier from these manifolds.

For example, consider a pair of saddle fixed points $a$ and $b$, 
of a two-dimensional map. 
A point $\eta \in \Ws(a) \cap \Wu(b)$ is a primary 
intersection point (p.i.p.) if (recall
$\Wu_\eta(b)$ is open and $\Ws_\eta(a)$ is closed)
\begin{equation}\label{eq:pip2D}
    \Ws_\eta(a) \cap \Wu_\eta(b) = \emptyset \;;
\end{equation}
that is, the manifolds up to $\eta$ intersect only at their boundary, \cite{Wiggins92}.  For
example, the points $p$ and $q$ are primary homoclinic points in
\Fig{fig:2DResonanceStd}, but the point $r$ is not since the set $\Ws_r(a) \cap
\Wu_r(b)$ contains five points. This definition will not work in higher
dimensions, since the  components of the set $\Ws(A) \cap \Wu(B)$ need not be
proper boundaries even when the intersections are transverse. Instead, we
generalize as follows:

\begin{defi}[Primary Intersection]\label{def:pip}
A point $\eta \in \Ws(A) \cap \Wu(B)$ is a \emph{primary intersection point}
if there exist proper boundaries $\gamma$ and $\sigma$ such that
\begin{equation}\label{eq:pip}
    \eta \in \Ws_{\sigma}(A) \cap \Wu_{f(\gamma)}(B) \;,\mbox{  but  } \;
    \Ws_{\sigma}(A) \cap \Wu_{\gamma}(B)= \emptyset \;.
\end{equation}
The set of primary intersections is denoted $\cP(A,B)$; it is an invariant set.

\end{defi}

\noindent For one-dimensional manifolds, this gives the same set as \Eq{eq:pip2D}.
Primary intersections can also be defined in terms of fundamental domains. Indeed,
since $\Ws_{\sigma} = \cS_{\sigma} \cup \Ws_{f(\sigma)}$ and
$\Wu_{f(\gamma)} = \cU_{\gamma} \cup \Wu_{\gamma}$ by \Eq{eq:decomp}, then if
$\eta$ is a primary intersection
\[
    \eta \in    \big( \cS_\sigma(A) \cup \Ws_{f(\sigma)}(A) \big)
           \cap \big( \cU_{\gamma}(B) \cup \Wu_{\gamma}(B) \big)
         = \cS_\sigma(A) \cap \cU_{\gamma}(B) \;,
\]
since, by \Eq{eq:pip} $\cS_{\sigma}(A) \cap \Wu_{\gamma}(B) = \cU_\gamma(B) \cap
\Ws_{f(\sigma)}(A) = \emptyset$. In addition, \Eq{eq:decomp} and \Eq{eq:pip}
imply that all forward images of the stable fundamental domain are disjoint from
the unstable one: $\cS_{f^t(\sigma)}(A) \cap \cU_\gamma(B) = \emptyset$ for $t >
0$. This means that the intersection index,
\begin{equation}\label{eq:index}
    \kappa( \cS, \cU) \equiv \sup \{t \in \Zset : f^t(\cS) \cap \cU \neq \emptyset \} \;,
\end{equation}
is zero. Consequently, an alternative characterization of the primary intersection
set is \cite{Lomeli00a}
\[
    \cP(A,B) = \{ \cS \cap \cU : \kappa(\cS, \cU) = 0,
        \cS \in \cF^\st(A), \cU \in \cF^\un(B) \} \;.
\]

We will henceforth make the assumption that the codimension-one manifolds
$\Ws(A)$ and $\Wu(B)$ have transversal primary intersections (the manifolds
may still have tangencies elsewhere). In this case, the components of $\cP(A,B)$
are codimension-two submanifolds that never cross. For example, in the
three-dimensional case, these components must be closed loops or
else are curves that are asymptotic to the invariant sets $A$ and $B$.

If the heteroclinic intersections appear as the result of splitting of
separatrices, then typically it is possible to apply Melnikov's method. Recall
that there is a correspondence between the zeroes of the Melnikov function and
heteroclinic intersections. If all the zeroes are simple then they continue precisely to 
the set of primary intersections.

\begin{figure}[th]
\begin{center}
\includegraphics[height=3.1in]{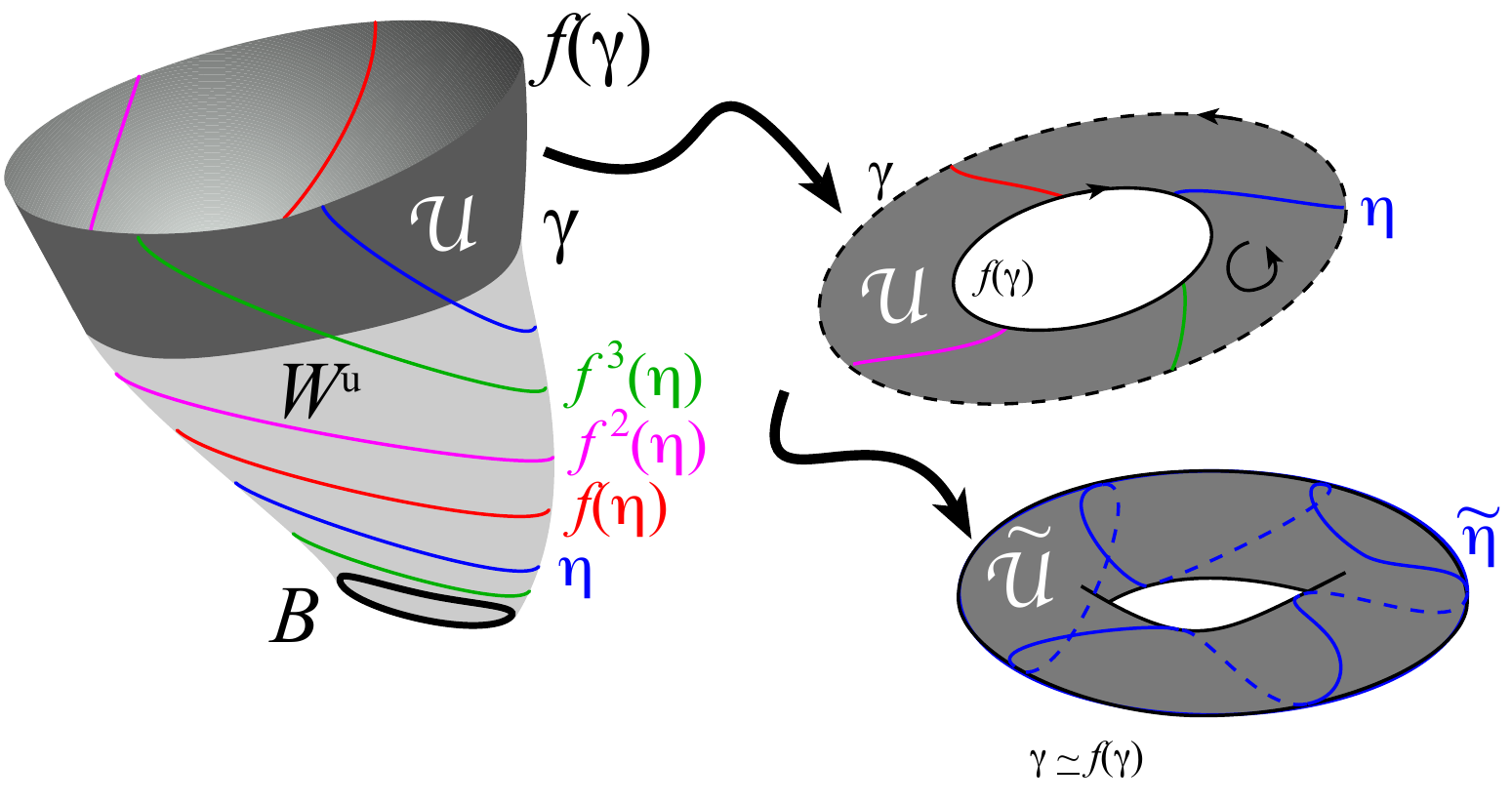}
\caption{Identifying the boundaries of a fundamental domain gives a torus
$\tilde\cU$ and primary intersection loop $\tilde\eta$.} \label{fig:fundamental}
\end{center}
\end{figure}

Since each fundamental domain generates the entire manifold, we can restrict
attention to the set of primary intersections in a particular fundamental domain,
say $\cP \cap \cU_\gamma$. A fundamental domain on a two-dimensional manifold is
an annulus, but under the natural identification $\gamma  \simeq f(\gamma)$ of its
boundaries, it can be thought of as a torus $\tilde\cU_\gamma$, see
\Fig{fig:fundamental}

If $\eta$ is a component of $\cP$ then, under this identification, its orbit
$f^t(\eta)$ is equivalent to a closed loop. Indeed, if $\eta$ is contained in the
interior of $\close{\cU_\gamma}$, then it must be a loop since the intersections
are assumed transverse. On the other hand, if there is an intersection point $p =
\eta \cap \gamma$, then since $\cP$ is invariant, the curve $f(\eta) \in \cP$ and
intersects $f(\gamma)$ at the point $f(p)$. Thus the components $\eta$ and
$f(\eta)$ are joined by the natural identification. 

Continuing this implies that the full
orbit of $p$ lies on $\gamma$ under the identification. This orbit must be finite,
since it would otherwise have limit points, violating transversality. Thus, the
orbit of $\eta$ becomes a closed loop $\tilde\eta$ on $\tilde\cU_\gamma$.

Therefore, on the torus $\tilde\cU_\gamma$, a primary intersection loop can be
labeled by its homology class, $(m,n) \in \Zset^2$. A loop that is homotopic to
$\gamma$ will be said to have class $(0,1)$, while loops that correspond to
intersections $\eta$ that are asymptotic to $A$ and $B$ will have class $(m,n)$
with $m \neq 0$. We previously used this classification to discuss bifurcations of
$\cP$ that occur when the manifolds develop tangencies as a parameter is varied
\cite{Lomeli00a, Lomeli03, Lomeli08d}.

\subsection{Resonance Zones}\label{sec:ResonanceZones}

In this section we will establish the basic assumptions to construct
resonance zones in terms of codimension one stable and unstable
manifolds of normally hyperbolic invariant manifolds.

The geometry of resonance zones based on arbitrary normally
hyperbolic invariant sets could be quite complicated. In order to gain some intuition, we start by describing  the case that $f$ is a map on a three-dimensional manifold $M$ and assume that $A$ and $B$ are hyperbolic fixed points or invariant circles. Motivated by this discussion, we will then propose a set of geometrical assumptions to define a partial barrier for a resonance zone.

\begin{figure}[th]
\begin{center}
\includegraphics[width=6in]{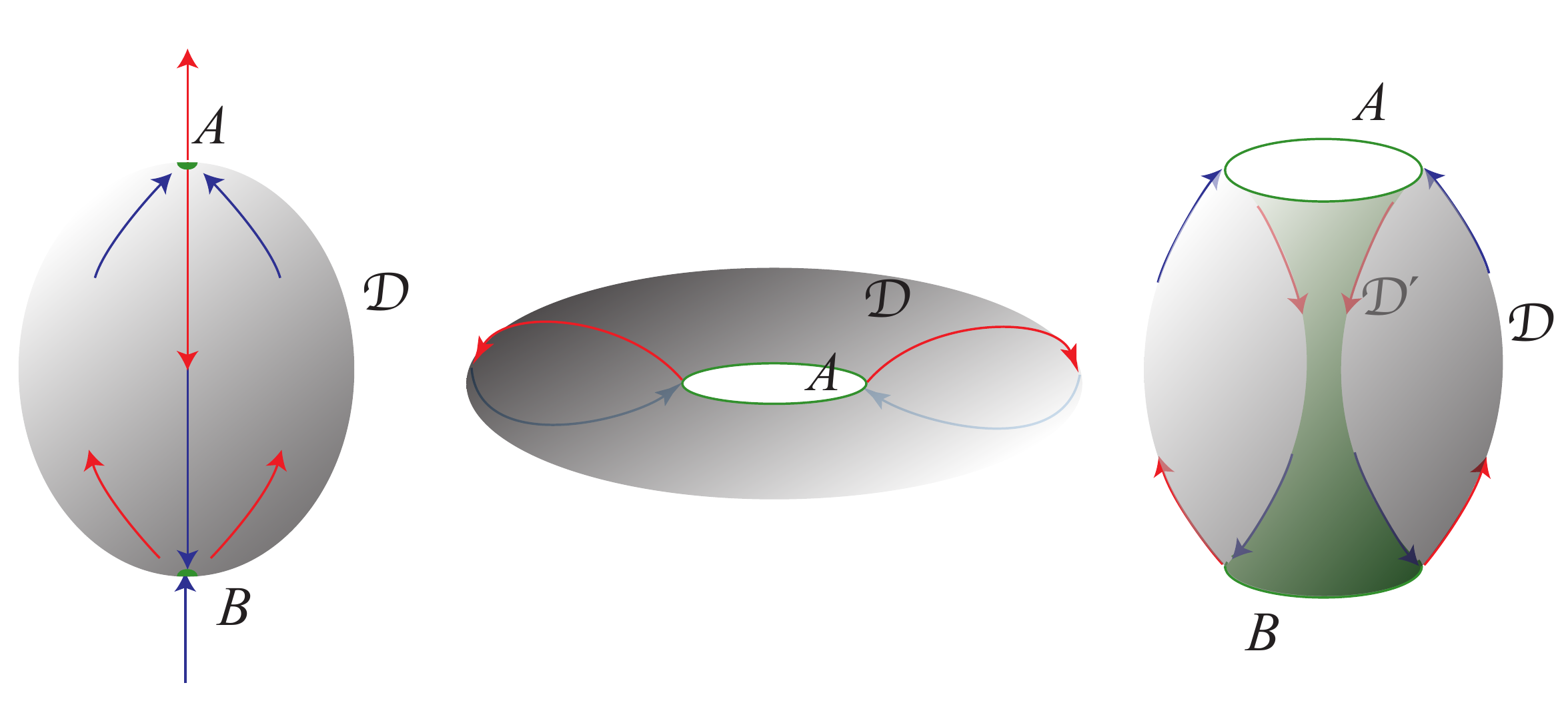}
\caption{Resonance zones in the integrable approximation for a pair of fixed
points, a single invariant circle or a pair of invariant circles.}
\label{fig:resonanceZones}
\end{center}
\end{figure}

Three typical resonance zones $\cR$ for this case are sketched in
\Fig{fig:resonanceZones}. For example, if $A$ and $B$ are fixed points, $\cR$ is a
ball that is, roughly speaking, bounded by $\Ws_\sigma(A)$ and $\Wu_\gamma(B)$
for some proper boundaries $\sigma$ and $\gamma$. The sketch corresponds to the
``integrable'' case when these manifolds coincide forming a saddle connection from
$A$ to $B$. More generally the boundary of $\cR$ will be made from a partial
barrier $\cD$ that is constructed from $\Ws_\sigma(A)$ and $\Wu_\gamma(B)$
plus a ``cap'' (see below).

When $A = B$ is a hyperbolic invariant circle and $\Wu(A)\cap\Ws(A) \neq
\emptyset$, the resonance zone is a solid torus. This is also true when $A \neq
B$, however, the boundary in this case is obtained from a pair of partial
barriers, $\cD$ constructed from $\Ws(A)$ and $\Wu(B)$ and $\cD'$ constructed from
$\Wu(A)$ and $\Ws(B)$ (which we also assume intersect); as before the simple case
where the partial barriers are saddle connections is sketched in
\Fig{fig:resonanceZones}. If $A$ is a fixed point and $B$ is an invariant circle,
then $\cR$ will typically be a ball, but a second fixed point or invariant circle
will be needed to complete the resonance, and its boundary will consist of at
least three partial barriers.

In the general case we require the following geometrical hypotheses.
\begin{enumerate}
\renewcommand{\theenumi}{\arabic{enumi}}
\renewcommand{\labelenumi}{\bf{(H\theenumi)}}
\setlength{\itemindent}{1cm}
    \item $A$ and $B$ are normally hyperbolic  invariant manifolds of 
    dimension at most $\frac{n-1}{2}$.;
    \item $\Ws(A)$ and $\Wu(B)$ are orientable, codimension-one submanifolds; and
    \item the set of primary intersections $\cP(A,B) \subset \Ws(A) \cap \Wu(B)$ is transverse.
\end{enumerate}

\begin{figure}
[th]
\begin{center}
\includegraphics[height=4in]{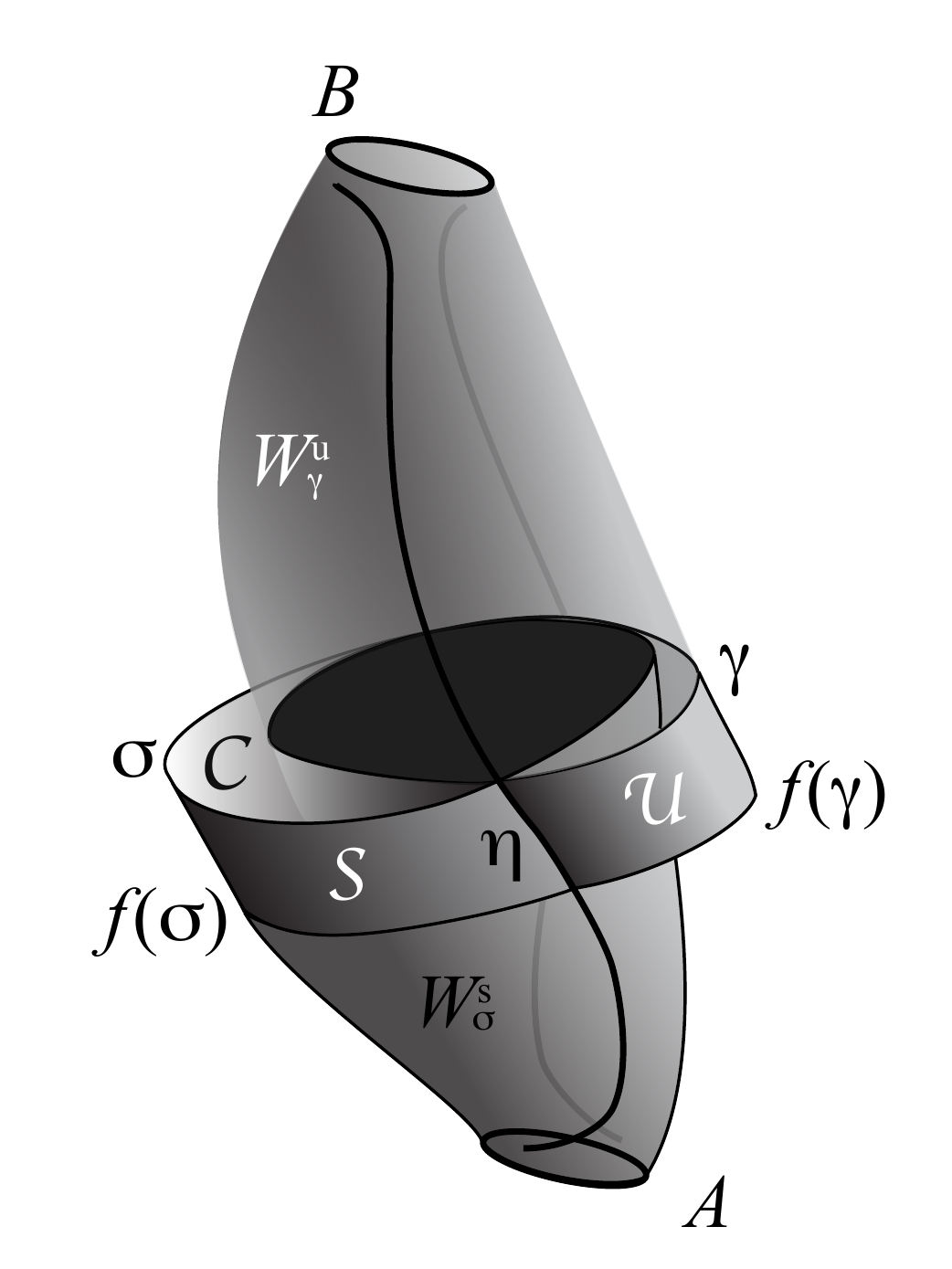}
\caption{Partial barrier constructed from $\Ws_\sigma(A)$, $\Wu_\gamma(B)$,
and a cap $\cC$.} \label{fig:intersections}
\end{center}
\end{figure}

To construct partial barrier $\cD$, begin by selecting a pair of proper boundaries
$\sigma$ and $\gamma$ such that the associated fundamental domains $\cS =
\cS_\sigma(A)$ and $\cU = \cU_\gamma(B)$ are in ``standard position'', i.e., such
that
\begin{enumerate}
\renewcommand{\theenumi}{\arabic{enumi}}
\renewcommand{\labelenumi}{\bf{(H\theenumi)}}
\setlength{\itemindent}{1cm} \setcounter{enumi}{3}
    \item $\kappa(\cS,\cU) = 0$ and the set $\eta \equiv \cS \cap \cU \subset \cP$  
a neat\footnote
{
      A submanifold with boundary $V \subset W$ is \emph{neat} if it is closed
      in $W$ and its boundary $\partial V$ is contained in the boundary $\partial W$ of $W$.
}
 submanifold of $\cS$
and $\cU$:\ \begin{equation}\label{eq:neatEta}
        \partial \eta \subset  \partial \cS  \cap \partial \cU \;.
    \end{equation}
\end{enumerate}
This condition is sketched in \Fig{fig:intersections}.

The simplest case corresponds to the primary intersection set containing a proper
boundary, for then we can choose $\gamma = \sigma \subset \cP$. In this case
\Hyp{4} is automatically satisfied since $\partial (\cS \cap \cU) = \emptyset$.
Recall that this is what is typically done for the two-dimensional case, as shown
in \Fig{fig:2DResonanceStd}. However $\cP$ does not always contain a proper
boundary, i.e., a curve with homology $(0,1)$; in particular, we commonly observe
that $\cP$ contains families of curves that spiral asymptotically from $B$ to $A$,
as sketched in \Fig{fig:fundamental} \cite{Lomeli00a, Lomeli03}. This also must
occur when the map is a Poincar\'e map of an autonomous flow, since heteroclinic
points of the map lie on heteroclinic orbits of the flow \cite{MacKay94}.

When resonance zone cannot have a boundary that consists solely of pieces of
stable and unstable manifolds we must add a \emph{cap} $\cC$ to construct a
partial barrier,
\begin{equation}\label{eq:partialBarrier}
    \cD = \Ws_\sigma(A) \cup \Wu_\gamma(B) \cup \cC \;,
\end{equation}
that will be one of the boundaries of $\cR$, see \Fig{fig:intersections}. We
assume that is it possible to choose the cap $\cC$ so that
\begin{enumerate}
\renewcommand{\theenumi}{\arabic{enumi}}
\renewcommand{\labelenumi}{\bf{(H\theenumi)}}
\setlength{\itemindent}{1cm} \setcounter{enumi}{4}
    \item $\cC$ is a codimension-one submanifold with boundary
        \begin{equation}\label{eq:cap}
            \partial \cC = \gamma \cup \sigma \;;
        \end{equation}
    \item
        $\cC \cap \Ws_\sigma(A) = \sigma$ and
        $\cC \cap \close{\Wu_\gamma(B)} = \gamma$.
\end{enumerate}

For the three-dimensional case the partial barrier under these assumptions 
is topologically a sphere when $A$ and $B$ are fixed points, 
like that sketched in \Fig{fig:resonanceZones}. When
$A=B$ is an invariant circle then $\cD$ is a torus, and when $A \neq B$ are
invariant circles, $\cD$ is an annulus bounded by the circles.

In the exceptional case that the set of primary intersections includes a proper
boundary, we choose $\sigma = \gamma \subset \cP$. In this case, $\cC =
\emptyset$, and the partial barrier is still given by \Eq{eq:partialBarrier}.

\section{Lobe Volume}\label{sec:lobeVolume}

 If the map is exact area-preserving, the lobe volume depends only upon the 
orbit of the manifolds, and can be computed using the
 generator $S$ of \Eq{eq:exactSymplectic} \cite{MMP84,MMP87, Easton91}. 
In this section we will show that the computation of lobe
 volume for the exact volume-preserving case will reduce to the integral 
of the one-form $\lambda$ of \Eq{eq:exact} along the
primary intersection curves.

Before proceeding, in \Sec{sec:formula}, to obtain the formula for lobe volumes,
we first derive an iterative formula relating the surface integral of $\alpha$
over a submanifold to those over its images.

\subsection{Iterative Formula}\label{sec:iterative}
\begin{figure}[th]
\begin{center}
\includegraphics[height=1.7in]{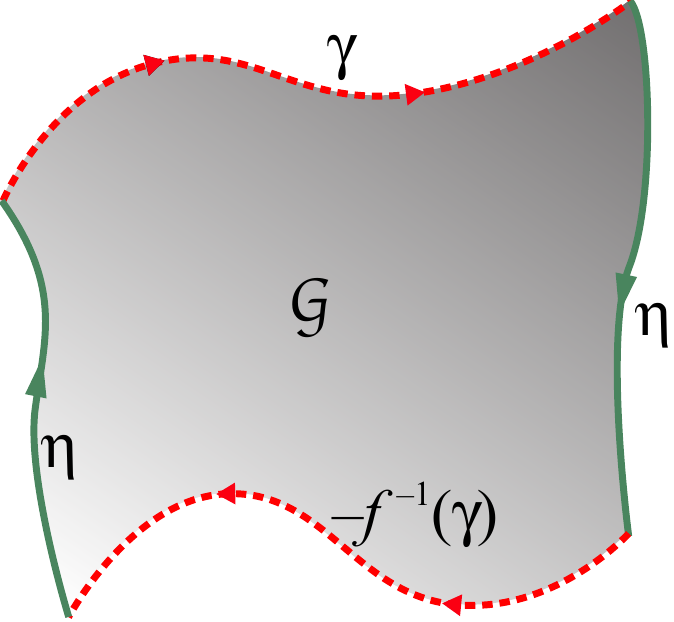}
\caption{$\cG$ is a submanifold such that a piece $\gamma$ of its boundary is
related through $f$ with another piece with opposite orientation,
$-f^{-1}(\gamma)$. In the diagram, an orientation of $\cG$ induces an orientation
on the boundary, and so $\partial\cG=\gamma-f^{-1}(\gamma)+\eta$.}
\label{fig:forlemma}
\end{center}
\end{figure}

Here we obtain a fundamental iteration formula for the integral of $\alpha$ over a
codimension-one submanifold $\cG$ that is arbitrary, except that its boundary can
be written as $\partial \cG = \gamma \cup f(\gamma) \cup \eta$, i.e., as the union
of two types of pieces: a manifold $\gamma$ and its image, and the remaining part
$\eta$ that is not related by iteration. This situation is sketched in
\Fig{fig:forlemma}.

\begin{lem}\label{lem:aux}
Suppose $f:M\to M$ is an exact volume-preserving diffeomorphism, \Eq{eq:exact},
and $\cG$ is an oriented codimension-one submanifold with boundary
 $\partial \cG=\gamma -f(\gamma)+\eta$, where $\gamma$ and $\eta$ are 
codimension-two submanifolds.
Then, for any $t \in \Nset$
\begin{equation}\label{eq:integralform}
\begin{split}
    \int_{\cG}\alpha + \int_{\gamma}\lambda
      &=    - \sum_{k=0}^{t-1}\int_{\eta} (f^k)^* \lambda +
      \int_{f^t(\cG)} \alpha + \int_{f^t(\gamma)} \lambda \; \\
      &=  \sum_{k=-1}^{-t}\int_{\eta} (f^k)^* \lambda +
      \int_{f^{-t}(\cG)} \alpha + \int_{f^{-t}(\gamma)} \lambda \;.
\end{split}\end{equation}
\end{lem}

\proof The composition formula \Eq{eq:composition} implies that
\[
     (f^t)^* \alpha - \alpha = \sum_{k=0}^{t-1} (f^k)^* d\lambda \;.
\]
Integration of this relation over the region $\cG$ and applying Stokes' theorem
gives
\[
    \int_\cG (f^t)^* \alpha - \int_{\cG} \alpha =
         \sum_{k=0}^{t-1} \int_{\eta +\gamma -f(\gamma)}(f^k)^* \lambda \;.
\]
The integrals over the images of the boundary curves $\gamma$ are a telescoping
sum, so that
\[
    \int_\cG (f^t)^* \alpha - \int_{\cG} \alpha =
         \sum_{k=0}^{t-1} \int_{\eta}(f^k)^* \lambda  +
         \int_{\gamma} \lambda -\int_{\gamma}(f^t)^* \lambda \;.
\]
Upon noting that, for example,
\[
    \int_{\cG}\left(f^t\right)^*\alpha=\int_{f^t(\cG)}\alpha \;,
\]
we see that the equation above is just the first line of \Eq{eq:integralform}
rearranged.

The remaining result can be obtained by a similar iteration, but backwards. Note
that \Eq{eq:exact} implies $\alpha - (f^{-1})^*\alpha = d ((f^{-1})^*\lambda)$;
this can be iterated to give
\[
     \alpha - (f^{-t})^*\alpha = \sum_{k=-1}^{-t} (f^k)^* d\lambda \;.
\]
Integrating this relation over $\cG$, as before, and rearranging gives the final
line of \Eq{eq:integralform}. \qed

\subsection{Lobe Volume Formula}\label{sec:formula}

We continue to assume that $f$ obeys \Hyp{1}-\Hyp{3} and has a partial barrier
$\cD$, \Eq{eq:partialBarrier}, constructed from fundamental domains $\cS$ and $\cU$ and a
cap $\cC$ that obey \Hyp{4}-\Hyp{6}. The turnstile for $\cD$ is the union of
the exit and incoming sets associated with the barrier; it is bounded by the
fundamental domains, $\cS$ and $\cU$, and the cap $\cC$ and its image $f(\cC)$
\[
    \partial(\cE \cup \cI) =  \cU \cup \cS \cup \cC \cup f(\cC) \;.
\]
Though $\cC$ is somewhat arbitrary, the lobe volume will be independent of this
choice because its boundary contains both $\cC$ and its image.

Since $\Ws(A)$ and $\Wu(B)$ have codimension one, in general they separate
the manifold $M$. A consistent orientation for $\Ws(A)$ and $\Wu(B)$ will define an ``outside'' and an ``inside'' of the barrier $\cD$. This in turn will induce an orientation of
$\sigma$ and, by iteration, an orientation of $f(\sigma)$. In this case we can
write $\partial\cS =\sigma- f(\sigma)$ and $\partial\cU =\gamma- f(\gamma)$.

\begin{figure}
[th]
\begin{center}
\includegraphics[height=2in]{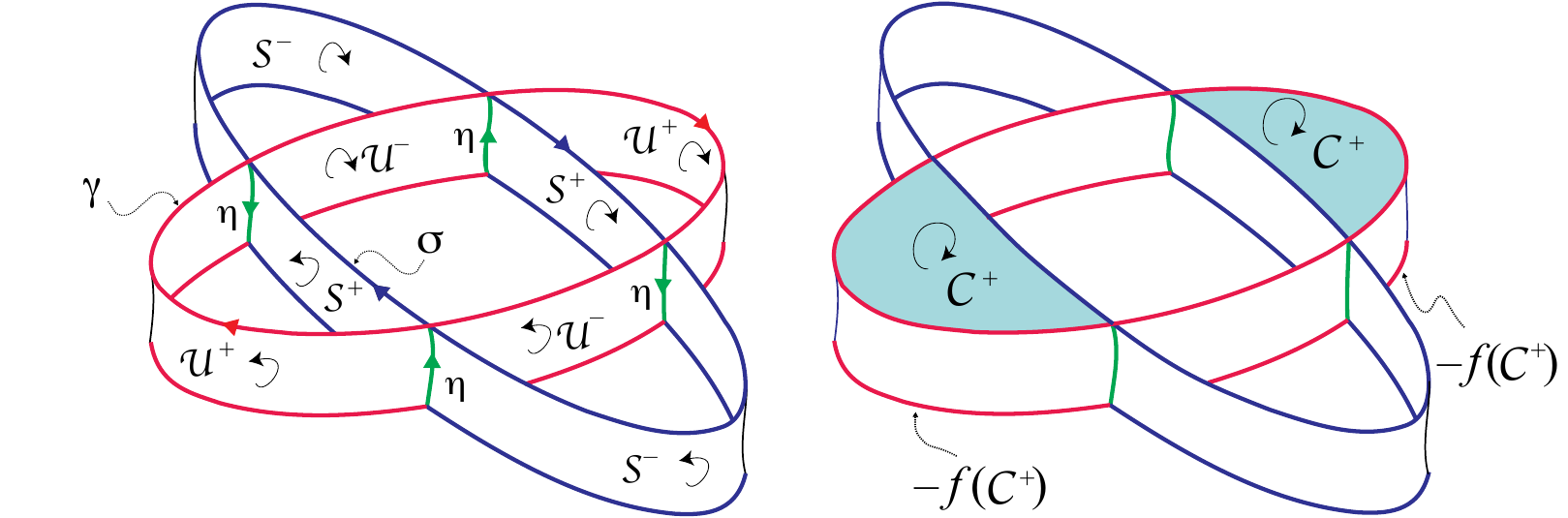}
\caption{Three-dimensional lobes. On the left is illustrated the boundary of the
exit and incoming lobes formed from a pair of fundamental domains that are subsets
of stable and unstable manifolds. On the right a pair of caps are added to
complete the exit lobe.} \label{fig:lobesGeneral}
\end{center}
\end{figure}

The exit set is the portion of the turnstile where the unstable manifold is
outside the stable manifold, we will use a $+$ sign to denote this subset. The
dividing set between exit and incoming lobes is the primary intersection $\eta =
\cS^+ \cap \cS^- = \cU^+ \cap \cU^- \subset \cP(A,B)$, which we have assumed is a
submanifold obeying \Eq{eq:neatEta}. Taking into account the orientation we write
\[
    \cS = \cS^+ + \cS^- \mbox{ and }
    \cU = \cU^+ + \cU^- \;,
\]
see \Fig{fig:lobesGeneral}. Thus the exit lobe has boundary
\begin{equation}\label{eq:exitLobe}
    \partial \cE =  \cU^+ - \cS^+  + \cC^+ - f(\cC^+) \;.
\end{equation}

The primary intersection $\eta$ also divides the loops $\sigma$ and $\gamma$ into
pieces that can be labeled $\pm$
\[
    \sigma=\sigma^+ + \sigma^- \mbox{ and } \gamma = \gamma^+ + \gamma^- \;,
\]
as sketched in \Fig{fig:lobesGeneral}. Consequently, the fundamental domains and
the cap have boundaries
\begin{equation}\label{eq:boundaries}
\begin{split}
    \partial\cS^\pm & = \pm \eta + \sigma^\pm -f(\sigma^\pm) \;, \\
    \partial \cU^\pm &= \pm \eta + \gamma^\pm -f(\gamma^\pm) \;, \\
    \partial \cC^\pm &= \sigma^\pm - \gamma^\pm \;.
\end{split}\end{equation}

We will compute the volume of the exit lobe \Eq{eq:exitLobe}, but as we discussed
in \Sec{sec:exitSet}, the volume of the incoming lobe is the same.

\begin{teo}[Lobe Volume]\label{thm:main}
Suppose that $\cD$ is a partial barrier for an exact volume-preserving map $f$,
obeying \Hyp{1}-\Hyp{6} and such that the one-form $\lambda$ is \emph{regular} on $\Ws(A)$ and $\Wu(B)$. Then the volume of the exit lobe $\cE$ with the boundary \Eq{eq:exitLobe} is
\begin{equation}\label{eq:lobeVolume}
    \volume{\cE} =\sum_{k\in\mathbb{Z} }\int_\eta (f^k)^*\lambda \;.
\end{equation}
\end{teo}
\proof

 If $\cS^+$ and $\cU^+$ are pieces of fundamental domains with boundaries obeying \Eq{eq:boundaries}, then Stokes' theorem with \Eq{eq:exact}, \Eq{eq:exitLobe}, and $\partial \cC^+ = \sigma^+ -\gamma^+$ gives
\begin{equation}\label{eq:lobeV1}
\begin{split}
     \int_\cE\volform{M}=\int_{\partial\cE}\alpha
        &= \int_{\cU^+}\alpha -\int_{\cS^+}\alpha +\int_{\cC^+} \alpha-\int_{f(\cC^+)}\alpha\\
        &= \int_{\cU^+}\alpha -\int_{\cS^+}\alpha - \int_{\partial\cC^+}\lambda\\
        &= \int_{\cU^+}\alpha+\int_{\gamma^+}\lambda
             -\int_{\cS^+}\alpha-\int_{\sigma^+}\lambda \;.
\end{split}
\end{equation}
By \Eq{eq:boundaries}, \Lem{lem:aux} applies to the terms in the last line of
\Eq{eq:lobeV1}, so that for all $t\in\Nset$,
\begin{equation}\label{eq:terms}
\begin{split}
    \int_{\cU^+}\alpha+\int_{\gamma^+}\lambda
       = & \sum_{k=-1}^{-t}\int_{\eta} (f^{k})^*\lambda 
         + \int_{f^{-t}(\cU^+)} \alpha
         + \int_{f^{-t}(\gamma^+)} \lambda \;,\\
    \int_{\cS^+}\alpha+\int_{\sigma^+}\lambda
      = -&\sum_{k=0}^{t-1}\int_{\eta} (f^k)^*\lambda 
        + \int_{f^t(\cS^+)} \alpha
        + \int_{f^t(\sigma^+)} \lambda \;.
\end{split}
\end{equation}
Here we have selected the direction of iteration to take advantage of the
contraction of the manifolds. Since by \Hyp{1}, $\dim(A)$ and $\dim(B)$ are at most $\frac{n-1}{2}$, \Lem{lem:aux1} implies that the $(n-1)$-form $\alpha$ is regular on $\Ws(A)$ and $\Wu(B)$. Thus, since $\close{\cU^+}\subset \Wu(B)$ and $\close{\cS^+}\subset \Ws(A)$ are compact \Lem{lem:aux2} implies
\[
        \lim_{t\to\infty}\int_{f^{-t}(\cU^+)} \alpha=0 \;,\quad
        \lim_{t\to\infty}\int_{f^t(\cS^+)}\alpha=0  \;.
\]
Finally, since $\lambda$ is assumed to be regular on $\Ws(A)$ and $\Ws(B)$, then
\begin{equation}\label{eq:lambdaLimits}
	\lim_{t\to\infty} \int_{f^{-t}(\gamma^+)} \lambda = 0 \;,  \quad
	\lim_{t\to\infty} \int_{f^t(\sigma^+)} \lambda  = 0 \;.
\end{equation}
After taking the limit in \Eq{eq:terms} and substituting back into \Eq{eq:lobeV1}, we find
\begin{align*}
  \int_\cE\volform{M}
    &= \sum_{k=-1}^{-\infty}\int_{\eta}(f^{k})^*\lambda +
       \sum_{k=0}^{\infty}\int_{\eta}(f^k)^*\lambda \;, 
\end{align*}
which is equivalent to \Eq{eq:lobeVolume}. This concludes the proof.\qed

\begin{remark}[1]
The primary intersection between $A$ and $B$ is given by
\[
	\cP(A,B)=\bigcup_{k\in\mathbb{Z} } f^k(\eta) \;.
\]
Therefore, under the assumptions of \Thm{thm:main}, \Eq{eq:lobeVolume} can be written
\[
	\int_\cE\volform{M}=\sum_{k\in\mathbb{Z} }\int_\eta (f^k)^*\lambda
	                   = \int_{\cP(A,B)}\lambda \;.
\]
\end{remark}

\begin{remark}[2]
By \Lem{lem:aux1}, the $(n-2)$-form $\lambda$ is regular on the $(n-1)$-dimensional manifolds $\Ws(A)$ and $\Ws(B)$ whenever $\dim(A), \dim(B)  \le \frac{n}{2}-1$. This includes the case that $A$ and $B$ are fixed points and $n \ge 3$; however, $\lambda$ is not necessarily regular for $n=3$ when $A$ or $B$ are invariant circles. When $A = B$, the hypothesis that $\lambda$ is regular is probably not needed. It is only the difference
\[
	\int_{f^{-t}(\gamma^+)} \lambda - \int_{f^t(\sigma^+)} \lambda
\] 
that must limit to zero in \Eq{eq:terms} to obtain \Eq{eq:lobeVolume}. This difference converges exponentially to zero if $f^{-t}(\gamma^+) \to A$ and $f^t(\sigma^+) \to A$ as $t \to \infty$ so that the domains of integration become identical. This is what happens for $n=2$ where $\lambda$ is the zero-form $S$, \Eq{eq:exactSymplectic}, \cite{MMP84, MMP87}, and for the examples in \Sec{sec:examples}. It would be nice to show that it is always possible to select proper boundaries so that this is the case. When $A \neq B$, it is not obvious whether the requirement that $\lambda$ be regular can always be satisfied. Since any closed form can be added to $\lambda$ without changing \Eq{eq:exact}, it may be possible to use this freedom to make $\lambda$ regular. One way to do this would be to have $\lambda|_A = \lambda|_B = 0$.
\end{remark}

\Thm{thm:main} implies that the volume of an $n$-dimensional exit set can be computed by
integrating along the $(n-2)$-dimensional primary intersection set. In general,
the volume of an $n$-dimensional lobe can be computed by integrating the 
$(n-2)-$form $\lambda$ on the primary intersection $\cP(A,B)$ that is an immersed
submanifold of dimension $n-2$. Moreover, the flux across a partial barrier 
determined by the manifolds $\Ws(A)$ and $\Wu(B)$ is independent of the selection 
of the fundamental domains and of the cap $\cC$.

Finally, note that \Thm{thm:main} also applies to the case that the cap $\cC$ is
the empty set. This would be the case if $\cP$ includes a proper loop, for then we
can select $\sigma = \gamma \in \cP$ and $\cD = \Ws_\sigma(A) \cup
\Wu_\sigma(B)$.

\section{Melnikov Flux}\label{sec:melnikov}

In this section, we show how our formula \Eq{eq:lobeVolume} for the lobe volume
limits to well-known Melnikov results for maps with a near saddle-connection. This
is well-known for the two-dimensional case \cite{MacKay88} and is implicit in the
theory developed in \cite{Lomeli00a} for the three-dimensional case. To prepare
for the result, we recall some notation and results of \cite{Lomeli03, Lomeli08a}.

Suppose $f_\delta:{M}\rightarrow{M}$ is a smooth family of exact volume-preserving
diffeomorphisms \Eq{eq:exact} satisfying \Hyp{1}-\Hyp{3} of
\Sec{sec:ResonanceZones} for each $0 < |\delta| < \delta_0$. In addition,
suppose that the map $f_0$ has a saddle connection
\[
    \Sigma \subset \Ws(A_0) \cap \Wu(B_0) \;.
\]
between the normally hyperbolic invariant manifolds $A_0$ and $B_0$ like those sketched in
\Fig{fig:resonanceZones}. As usual, we assume that $\dim(\Sigma)=n-1$ and
$\dim(A_0),\dim(B_0)\le \frac{n-1}{2}$.

By \Hyp{3}, when $\delta > 0$ the saddle connection splits into the stable and
unstable manifolds $\Ws(A_\delta) = \Ws(A_\delta,f_\delta)$ and $\Wu(B_\delta) = \Wu(B_\delta,f_\delta)$ of the perturbed invariant sets. The splitting of these manifolds is computed, to lowest order, by the classical Melnikov function.

Specifically, suppose there is a fundamental domain $\cF \subset \Sigma$ and two
diffeomorphisms $\psi^{s,u}_\delta:\cF\subset\Sigma\to M$  (adapted deformations), such that
\[
    \cU_\delta =\psi^\un_\delta (\cF) \mbox{ and } \cS_\delta =\psi^\st_\delta (\cF)
\]
are fundamental domains of the perturbed manifolds $\Ws(A_\delta)$ and
$\Wu(B_\delta)$. When $\delta = 0$, we can take $\psi^{\un,\st}_0 $ 
to be the trivial inclusion $\psi^{s,u}_\delta:\cF\subset\Sigma\to M$,
$\psi^{\un,\st}_0(\xi)=\xi $.
so that
\[
    \cU_0 = \cS_0 = \cF \;.
\]
We notice that the intersection index $\kappa(\cS_0,\cU_0) = \kappa(\cF,\cF) = 0$, since $\cF$ is
a fundamental domain, but this need not be true for $\cS_\delta$ and $\cU_\delta$,
unless the deformations were chosen carefully.
Nevertheless, as in \Hyp{3}, we assume that the set of primary intersection
$\eta_\delta$ is transverse for small $\delta$. As we will see, the limiting set
$\eta_0 \cap \cF$ is the zero set of a Melnikov function.

To first order in $\delta$ the only relevant quantities associated with $f_\delta$ and
the deformations $\psi^{\st,\un}_\delta$ are their vector fields with respect to
$\delta$. The perturbation of the map $f_0$ away from $\delta = 0$ is measured by its
\emph{perturbation vector field}
\begin{equation}\label{eq:pertvect}
      X(x) \equiv \left.\frac{\partial }{\partial \delta}\right|_{\delta = 0}
                   f_{\delta }(f_{0}^{-1}(x)) \;,
\end{equation}
for each $x \in M$. The perturbations of the deformations $\psi^{\st,\un}_\delta$
are similarly defined by the vector fields
\[
    Z^{\st,\un} (\xi) = \left. \frac{\partial}{\partial\delta}\right|_{\delta=0}
                      \psi^{\st,\un}_\delta(\xi) \;,
\]
for each $\xi \in \cF$, using the assumption that $\psi^{\st,\un}_0 = id$ on
$\cF$.

Up to vectors tangent to the saddle connection, these vector fields are related by
the iterative formulas \cite{Lomeli08a}
\begin{equation}\label{eq:Ziteration}
     (f_0)_* Z^{\un,\st} -Z^{\un,\st} + X_0 \in T\cF \;, \\
\end{equation}
\begin{equation}\label{eq:Ziteration2}
      Z^{\un,\st} -f^*_0 Z^{\un,\st} + f^*_0 X_0 \in T\cF \;. \\
\end{equation}
To see this, note that since $\Sigma$ is invariant under $f_0$, the functions
$\tilde \psi^{\st,\un}_{\delta} =f_\delta\circ \psi^{\st,\un}_{\delta }\circ f^{-1}_0 $
are also adapted deformations, mapping $\Sigma$ to $W^{\st,\un}_\delta$. In either
case, the function $C_\delta = \tilde \psi_\delta^{-1} \circ \psi_\delta$ defines a
curve in $\cF$, so that the vector $\frac{d}{d\delta} C_\delta$ is in $T\cF$.
Computing this derivative implies that
\[
     \left[
        \frac{\partial}{\partial\delta } \tilde \psi_{\delta }(x) -
        \frac{\partial}{\partial\delta }        \psi_{\delta }(x)
     \right]_{\delta =0}
     \in T_{\xi}\cF \;.
\]
However,
\[
     \left[
        \frac{\partial}{\partial\delta } \tilde \psi_{\delta }(x)
     \right]_{\delta =0} = X_0(\xi)+Df_0(f^{-1}_0(\xi))Z( f^{-1}_0(\xi)) \;.
\]
Recalling the definition of the pullback and pushfoward (see \Eq{eq:pullbackV}) and combining
these two results gives the pair of formulas \Eq{eq:Ziteration} and \Eq{eq:Ziteration2}.

Since the maps $\psi^{\st,\un}$ describe the deformations of the stable and
unstable manifolds, the ``velocity'' of the splitting with $\delta$ is given by the
deformation vector field
\begin{equation}\label{eq:delta}
    {\Delta}(\xi)=Z^\un(\xi)-Z^\st(\xi) \;.
\end{equation}
The stable and unstable manifolds are thus ``flowing'' with respect to $\delta$
according to the vector fields $Z^{\st,\un}$ and they split at a rate $\Delta$.
This flow induces a ``flux" with respect to the volume form $\volform{M}$ defined by
\begin{equation}\label{eq:fluxForm}
    \Phi \equiv i_\Delta \volform{M} \;.
\end{equation}
The form $\Phi$ is the flux form of the deformation $\Delta$.

Though the expression for the splitting velocity $\Delta$ depends on the choice of
adapted deformations, the form $i_\Delta\volform{M}$ is well-defined on $\Sigma$ and is
independent of the choice of $\psi^{s,u}$ \cite{Lomeli03}. In addition, from the
definition \Eq{eq:delta} with \Eq{eq:Ziteration} we have
\begin{equation}\label{eq:deltaInv}
    f^*_0{\Delta}- {\Delta} \in T\cF \;;
\end{equation}
which implies that
\[
    f_0^*\left( i_\Delta\volform{M}\right) =i_{f_0^*\Delta}\left(f_0^*\volform{M}\right)
                    =i_{f_0^*\Delta}\volform{M}=i_\Delta\volform{M} \; \mbox{ on }\; \Sigma \;,
\]
i.e.,  the flux is invariant under $f_0$.

The flux is our primary measure of the splitting; indeed as we will show below,
the integral of $\Phi$ is related to the lobe volume. Since $f_\delta$ is exact, the
net flux across a fundamental domain vanishes.

\begin{pro}[\cite{Lomeli03}] If $\cF$ is a fundamental domain in $\Sigma$ then
\[
    \int_\cF \Phi = 0 \;.
\]
\end{pro}

The classical Melnikov function is defined relative to a choice of an adapted
normal vector field $\bn: \Sigma \to TM$ \cite{Lomeli00a}. Recall that vector
field is \emph{adapted} to $f_0$ if, for all vector fields $Y$ and all points $\xi
\in \Sigma$,
\begin{equation} \label{eq:adaptVect}
     f_0^* \langle \bn,Y\rangle = \langle \bn,f_0^* Y\rangle \;.
\end{equation}
Given an inner product $\langle \;,\; \rangle$ and  such a vector field, the
Melnikov function $M_\bn: \Sigma \to \Rset$ is simply the normal component of the deformation,
\begin{equation}\label{eq:Melnikov}
    M_\bn = \langle \bn , \Delta \rangle \;.
\end{equation}
Since $\bn$ is a vector field in the algebraic normal bundle of $\Sigma$, $M_\bn$
is a measure of the speed of splitting in the normal direction. Conditions
 \Eq{eq:deltaInv}  and \Eq{eq:adaptVect}together imply that $M_\bn$ is invariant
under $f_0$, that is, $M_\bn\circ f_0=M_\bn$. One common choice for $\bn$ is the
gradient of an invariant of $f_0$ (if one exists) \cite{Lomeli00a}, but any
adapted normal vector field gives a Melnikov function with the same zero set.

The main point of Melnikov theory is that if a point $\xi \in \Sigma$ is a
nondegenerate zero of $M_\bn $, the stable and unstable manifolds $\Wu(B_\delta)$
and $\Ws(A_\delta)$ intersect transversally near $\xi$ when $\delta$ is small enough.
The flux form is simply related to the classical Melnikov formula by
\begin{equation}\label{eq:MelFlux}
    \Phi = M_\bn \omega_\bn \;,
\end{equation}
where $\omega_\bn$ is the natural $(n-1)$-volume form on $\Sigma$ induced by
$\bn$ \cite{Lomeli03, Lomeli08a},
\begin{equation}\label{eq:omegan}
    \omega_\bn=\frac{1}{\langle \bn,\bn\rangle}i_\bn\volform{M} \;.
\end{equation}
When $\bn$ is an adapted normal,  $\omega_\bn$ is invariant under $f_0$:
$f_0^*\omega_\bn=\omega_\bn$ on $\Sigma$.

The iterative formula \Eq{eq:Ziteration} implies \cite{Lomeli03,Lomeli08a}
\begin{equation}\label{eq:deltaSum}
       \Delta - \sum_{k=-\infty}^{\infty}\left( f^*_0\right)^{k}   X  \in T\cF  \;.
\end{equation}
Indeed, the function $\Delta$ can be thought of as a section on a
normal bundle. It was proven in \cite{Lomeli08a} that if one considers both $\Delta$ and the series $\sum_{k=-\infty}^{\infty}\left( f^*_0\right)^{k}  X $ as sections, then they are
the same function.

That is, the deformation vector field is, up to tangent vectors, given by the
infinite sum above. This infinite sum is geometrically convergent since the
invariant sets $A_0$ and $B_0$ are normally hyperbolic. By \Eq{eq:Melnikov} and
\Eq{eq:deltaSum} the Melnikov function can be written
\begin{equation}\label{eq:Mnsum}
   M_\bn = \sum_{k=-\infty}^{\infty}\langle \bn,\left( f^*_0\right)^{k} X\rangle=
           \sum_{k=-\infty}^{\infty}\langle \bn,  X\rangle\circ  f^k_0
\end{equation}
since, by definition, $\langle \bn, Y \rangle = 0$ for any $Y \in T\cF$.

There is considerable freedom in the selection of the adapted deformations
$\psi_\delta^{\st,\un}$; we may choose any functions from $\cF$ to $W^{\st,\un}$
that reduce to the identity when $\delta = 0$. Given an given adapted
normal $\bn$, there is a natural choice: for each $\xi \in \cF$ choose
$\tilde\psi^\st_\delta(\xi)$ ($\tilde\psi^\un_\delta(\xi))$ to belong to the
intersection of the stable (unstable) manifold with the line generated by
$\bn(\xi)$. Under this choice $\widetilde{\Delta}=\phi \bn$ for some function
$\phi:\Sigma\to\Rset$ to be determined.

Since $\tilde \Delta$ is also a deformation vector field, \Eq{eq:deltaSum} implies that
$\Delta-\widetilde{\Delta} \in T\cF$, so that $M_\bn = \langle \bn, \tilde \Delta
\rangle$, as well. From this, we conclude that $M_\bn = \langle \bn,\bn\rangle
\phi$ and therefore $\widetilde{\Delta}$ has to be of the form
\begin{equation}\label{eq:tilde}
 \widetilde{\Delta}=\left( \frac{M_\bn}{\langle \bn,\bn\rangle}\right) \bn.
\end{equation}

We are now prepared to state relate these old results to \Thm{thm:main} showing that
an integral of the flux form $\Phi$ gives the rate of growth of the lobe volume with $\delta$.

\begin{teo} Let $f_\delta$ be a $C^1$ family of exact volume-preserving diffeomorphisms
with partial barriers $\cD_\delta$ and exit sets $\cE_\delta$ obeying \Hyp{1}-\Hyp{6}, such that $f_0$ has a saddle connection
\[
    \Sigma \subset \Ws(A_0) \cap \Wu(B_0) \;.
\]
Lett $M_\bn$ be a Melnikov function \Eq{eq:Melnikov} for $f_\delta$ and
$\cF^+ =\cF\cap M_{\bn}^{-1}(\Rset^+)$ be the set of points in the fundamental domain $\cF$ for which the Melnikov function is positive. Then
\begin{equation}
   \left. \frac{d}{d \delta}\right|_{\delta = 0} \int_{\cE_\delta} \volform{M} 
	= \int_{\cF^+} \Phi 
	=\frac12\int_{\cF} |M_{\bn}|\omega_{\bn} \;.
\end{equation}
\end{teo}

\proof By assumption $f_\delta$ is exact-volume preserving so there exists a family
of $(n-2)$-forms $\lambda_\delta$ such that
\[
    f_\delta^* \alpha - \alpha = d\lambda_\delta \;.
\]
Note that since $f_\delta$ is assumed to be smooth with respect to $\delta$, we can
take $\lambda_\delta$ to be smooth as well.

Let $\partial \cF = \sigma_0 - f_0(\sigma_0)$ for the proper boundary $\sigma_0$
and define $\sigma_0^+=\sigma_0\cap \cF^+$ so that the boundary of $\cF^+$ is
of the form
\begin{equation}\label{eq:FplusBound}
    \partial {\cF^+}=\sigma_0^+-f_0(\sigma_0^+) +\eta_0 \;.
\end{equation}
According to \Eq{eq:lobeV1}
\begin{equation}\label{eq:prop61}
    \int_{\cE_\delta} \volform{M}
        = \int_{\cU^+_\delta }\alpha - \int_{\cS^+_\delta}\alpha
        + \int_{\gamma^+_\delta}\lambda_\delta - \int_{\sigma^+_\delta}\lambda_\delta \;.
\end{equation}
Using the adapted deformations $\psi^{\st,\un}_\delta$ to map $\cF^+$ to $\cS^+_\delta$ and
$\cU^+_\delta$, respectively, then gives
\begin{equation}\label{eq:initialformula}
 \int_{\cE_\delta} \volform{M}
    = \int_{\cF^+}\left[ (\psi^\un_\delta )^*\alpha -(\psi^\st_\delta)^*\alpha \right]
    + \int_{\sigma^+_0} \left[ (\psi^\un_\delta)^*\lambda_\delta-(\psi^\st_\delta)^*\lambda_\delta \right] \;.
\end{equation}
Differentiate this relation with respect to $\delta$ and use the definition \Eq{eq:delta} to obtain
\begin{equation}\label{eq:first}
     \left.\frac{\partial}{\partial\delta}\right|_{\delta=0} \int_{\cE_\delta} \volform{M}
     =   \int_{\cF^+} L_\Delta \alpha  + \int_{\sigma^+_0} L_\Delta \lambda_0 \;,
\end{equation}
where $L_\Delta$ is the Lie derivative \Eq{eq:LieDeriv}.

Cartan's formula \Eq{eq:LieIdentity} implies that
\begin{align*}
    L_\Delta \alpha &= i_\Delta d\alpha + d(i_\Delta \alpha)
            = i_\Delta \volform{M} + d(i_\Delta \alpha) \;,\\
    L_\Delta \lambda_0 &= i_\Delta d\lambda_0 + d(i_\Delta \lambda_0)
            = i_\Delta (f^*_0\alpha) - i_\Delta \alpha + d(i_\Delta \lambda_0) \;,
\end{align*}
where we used $d\alpha = \volform{M}$ and the exactness of $f_0$. Substitution of these
results into \Eq{eq:first} yields
\[
  \left.\frac{\partial}{\partial\delta}\right|_{\delta=0}\int_{\cE_\delta} 
\volform{M} = \int_{\cF^+} i_\Delta\volform{M}  + N \;,
\]
where
\begin{equation}\label{eq:Ndefn}
   N \equiv\int_{\partial \cF^+} i_\Delta \alpha
        +  \int_{\sigma^+_0} i_\Delta (f^*_0\alpha)
        - \int_{\sigma^+_0} i_\Delta \alpha + \int_{\partial \sigma^+_0} i_\Delta \lambda_0 \;.
\end{equation}
Consequently, in order  to prove the theorem we must finally show that $N = 0$.
Using \Eq{eq:FplusBound} for $\partial \cF^+$, then \Eq{eq:Ndefn} becomes
\begin{equation}\label{eq:Ntwo}
\begin{split}
    N &=  \int_{\sigma^+_0} \left[ i_\Delta(f_0^* \alpha) - f_0^*(i_\Delta \alpha) \right]
       + \int_{\eta_0} i_\Delta \alpha + \int_{\partial \sigma^+_0} i_\Delta \lambda_0  \\
      &= \int_{\sigma^+_0} i_{(\Delta-f_0^*\Delta)}(f_0^* \alpha)
       + \int_{\eta_0} i_\Delta \alpha + \int_{\partial \sigma^+_0} i_\Delta \lambda_0  \;.
\end{split}\end{equation}

Note that this expression for $N$ is independent of the choice of $\alpha$. In
order words, if $\widetilde{\alpha}$ is {\em any} form such that
$d\widetilde{\alpha}=\volform{M}$ and
$f^*_\delta\widetilde{\alpha}-\widetilde{\alpha}=d\widetilde{\lambda}_\delta$ then we
may begin again using the new forms in the computation of \Eq{eq:prop61},
obtaining finally \Eq{eq:Ntwo} with $\alpha \to \widetilde{\alpha}$ and $\lambda_0
\to \tilde\lambda_0$. In particular, if $f_0$ is exact with respect to
$\alpha$, then it is also exact with respect to $\widetilde{\alpha}=\left(
f^*_0\right)^{k-1}\alpha $, for all $k\in\Nset$. Using this new form in
\Eq{eq:Ntwo} gives
\[
  N = \int_{\sigma^+_0} i_{(\Delta-f_0^*\Delta)} (f_0^*)^{k} \alpha
       + \int_{\eta_0} i_\Delta (f_0^*)^{k} \alpha
       + \int_{\partial \sigma^+_0} i_\Delta (f_0^*)^{k}\lambda_0\;.
\]

Finally, note that $N$ is also independent of the choice of $\Delta$: since \Eq{eq:initialformula} is independent of the choice of adapted deformations, \Eq{eq:first} is also independent of the choice of $\Delta$---this equation and the subsequent ones are valid for any deformation $\tilde{\Delta}$ that comes from a pair of adapted deformations $\tilde{\psi}^{\st,\un}_\delta$. 
In particular, we will use the deformation \Eq{eq:tilde}, so that $\Delta \to \tilde \Delta$, which is normal to $\cF$ and thus $\tilde \Delta|_{\eta_0} = \tilde \Delta_{\sigma_0^+}= 0$ since $\partial \sigma_0^+ \subset \eta_0 \subset \cF$.
Thus if we define $Y=\tilde \Delta-f_0^*\tilde \Delta$ then $Y \in T\cF$ by \Eq{eq:deltaInv} and
\[
  N = \int_{\sigma^+_0} i_{Y} (f_0^*)^{k} \alpha\; .
\]
Since the $(n-1)$-form $\alpha$ is regular when $\dim(A_0), \dim(B_0) \le \frac{n-1}{2}$, as assumed in \Hyp{1}, \Lem{lem:aux2} implies that as $k\to\infty$, $N \to 0$; however, $N$ is independent of $k$, therefore $N \equiv 0$.
\qed

\section{Examples}\label{sec:examples}

Here we give two simple examples demonstrating that \Eq{eq:lobeVolume} gives the
expected results for the lobe volume. In the first case, the lobes are simply
given by the cross product of the lobes for an area preserving map with a circle.
In the second, the map is the time $T$ map of a nonautonomous flow.

\subsection{Semidirect Product with a Twist Map}
An area-preserving, twist map $(X,Y) = g(x,y)$ can be obtained from an implicit
generating function $G(x,X)$ through the equations
\begin{equation}\label{eq:symplecticGenerator}
    Y = \partial_2 G(x,X) \;, \quad y = -\partial_1 G(x,X) \;.
\end{equation}
provided that $G$ satisfies the \emph{twist} condition $\partial_1 \partial_2
G\neq 0$. A map generated in this way is also exact in the sense of
\Eq{eq:exactSymplectic}. Indeed we may choose $\nu = y dx$ and the zero-form
\begin{equation}\label{eq:LfromS}
    S(x,y) = G(x,X(x,y)) \;,
\end{equation}
for in this case
\[
    g^*\nu -\nu = Y dX - y dx \;,
\]
and
\[
dS(x,y) = \partial_1 G(x,X) dx + \partial_2 G(x,X)dX \;,
\]
which are equivalent to \Eq{eq:symplecticGenerator}.

The map $g$ can be extended to an exact volume-preserving map on $M=\Rset^2 \times
\Sset$ by introducing an angle $\theta \in \Sset \equiv \Rset/(2\pi \Zset)$, and some
appropriate dynamics. A simple case is the semidirect product
\begin{equation}\label{eq:semiD}
    (X,Y,\Theta) = f(x,y,\theta) = \left(g(x,y), \theta + \rho(x,y) \right) \;,
\end{equation}
where $\rho$ is the local rotation number.  Now $f$ is an exact volume-preserving
map with the volume form $\volform{M} = dx \wedge dy \wedge d\theta$. For example,
using the two form
\[
    \alpha =d\left(S\circ f^{-1}\right) \wedge d\theta -ydx \wedge d\theta \;,
\]
so that $\volform{M} = d\alpha$, and $dS = YdX - ydx$ gives
\begin{align*}
    f^* \alpha -\alpha
    &= dS\wedge d\Theta  -YdX \wedge d\Theta
     -d\left( S\circ f^{-1}\right) \wedge d\theta + ydx \wedge d\theta   \\
    &= - d\left( S\circ f^{-1}\right) \wedge d\theta - y \left(\partial_y \rho\right) \, dx\wedge dy \;.
\end{align*}
Therefore, $f$ obeys \Eq{eq:exact} with the one form
\begin{equation}\label{eq:semiForm}
    \lambda = -\left( S\circ f^{-1}\right) d\theta  -\phi(x,y)y\, dy.
\end{equation}
where $\phi $ is a function such that $\partial_x \phi = \partial_y \rho$.

Because of the semidirect product structure of \Eq{eq:semiD}, if  $a$ is a saddle fixed point of $g$, then $A = a \times \Sset$ is a normally hyperbolic invariant circle for $f$. Similarly
if $\cR$ is a resonance zone for $g$ (recall for example \Fig{fig:2DResonanceStd}) then $\bar\cR = \cR \times \Sset$ is a resonance zone for $f$. Moreover, if $\cE$ is the exit
set for $\cR$ that is delineated by a pair of primary intersection points $p$ and
$q$, then $\bar\cE = \cE \times \Sset$ is the exit set for the three-dimensional
resonance zone of $f$. Moreover, since for each $(x,y)$ the map $\Theta = \theta +
\rho$ is a homeomorphism of the circle $\Sset$, the primary intersection manifold of the
fundamental domains for $f$ is the submanifold
\begin{equation}\label{eq:primaryPQ}
    \eta = \eta_p -\eta_q
\end{equation}
where $\eta_p = \{p\} \times \Sset$ and $\eta_q = \{q\} \times \Sset$, and $p,q$
are in the primary homoclinic orbits for $g$, recall \Fig{fig:2DResonanceStd}.

Thus we can apply \Thm{thm:main} to compute the volume of $\cE \times \Sset$
\begin{align*}
    \volume{\bar\cE} = \sum_{k\in \mathbb{Z}}\int_{\eta_p -\eta_q}( f^{k})^* \lambda
    &= \sum_{k\in \mathbb{Z}}\left( \int_{f^k(\eta_p)}\lambda
      - \int_{f^k(\eta_q)}\lambda\right),
\end{align*}
where $\lambda$ is the one form \Eq{eq:semiForm}. These sums converge without the assumption that $\lambda$ is regular because $f^k(\eta_{p,q}) \to A$ as $k \to \pm \infty$.
Noting that $f^k(\eta_p) =
g^k(p) \times \Sset$, and that $dy$ vanishes on the orbits of the primary
intersections, the integrals above can be easily computed. For example,
\[
  \int_{f^k(\eta_p)}\lambda = -S(g^{k-1}(p)) \int_\Sset d\theta =  -2\pi S(g^{k-1}(p)) \;.
\]
Hence the volume becomes
\begin{equation}\label{eq:semiDVolume}
 \volume{\bar \cE} = 2\pi \sum_{k\in \mathbb{Z}} \left( S(g^k(q)) - S(g^k(p)) \right)\;.
\end{equation}
which  $2 \pi$ times the difference between the actions of the orbits of $q$ and
$p$ under the symplectic map $g$. The action difference is exactly the lobe area
for an area-preserving map as shown in \cite{MMP87,Easton91}, so that we have
derived the obvious formula
\[
    \volume{\bar \cE} = 2\pi \mbox{ area}(\cE) \;.
\]

As an explicit example consider the generator
\[
    G(x,X)=\frac{1}{2}\left( X-h(x)\right) ^{2} \;,
\]
for a diffeomorphism $h:\Rset \to \Rset$. This generator satisfies the twist
condition since $\partial_1 \partial_2 G = -h'(x) \neq 0$ by assumption. The
generated map is, explicitly, given by
\begin{equation}\label{eq:HermanMap}
    g(x,y) =(X,Y)= \left( h(x) + \frac{y}{h'(x)},\frac{y}{h'(x)} \right).
\end{equation}
For this case, the generator \Eq{eq:LfromS} is
\begin{equation}\label{eq:paupau}
    S(x,y) =  \frac12 \left(\frac{y}{h'(x)}\right)^2 =\frac12 Y^2 \;.
\end{equation}

\begin{figure}[th]
\begin{center}
\includegraphics[height=3in]{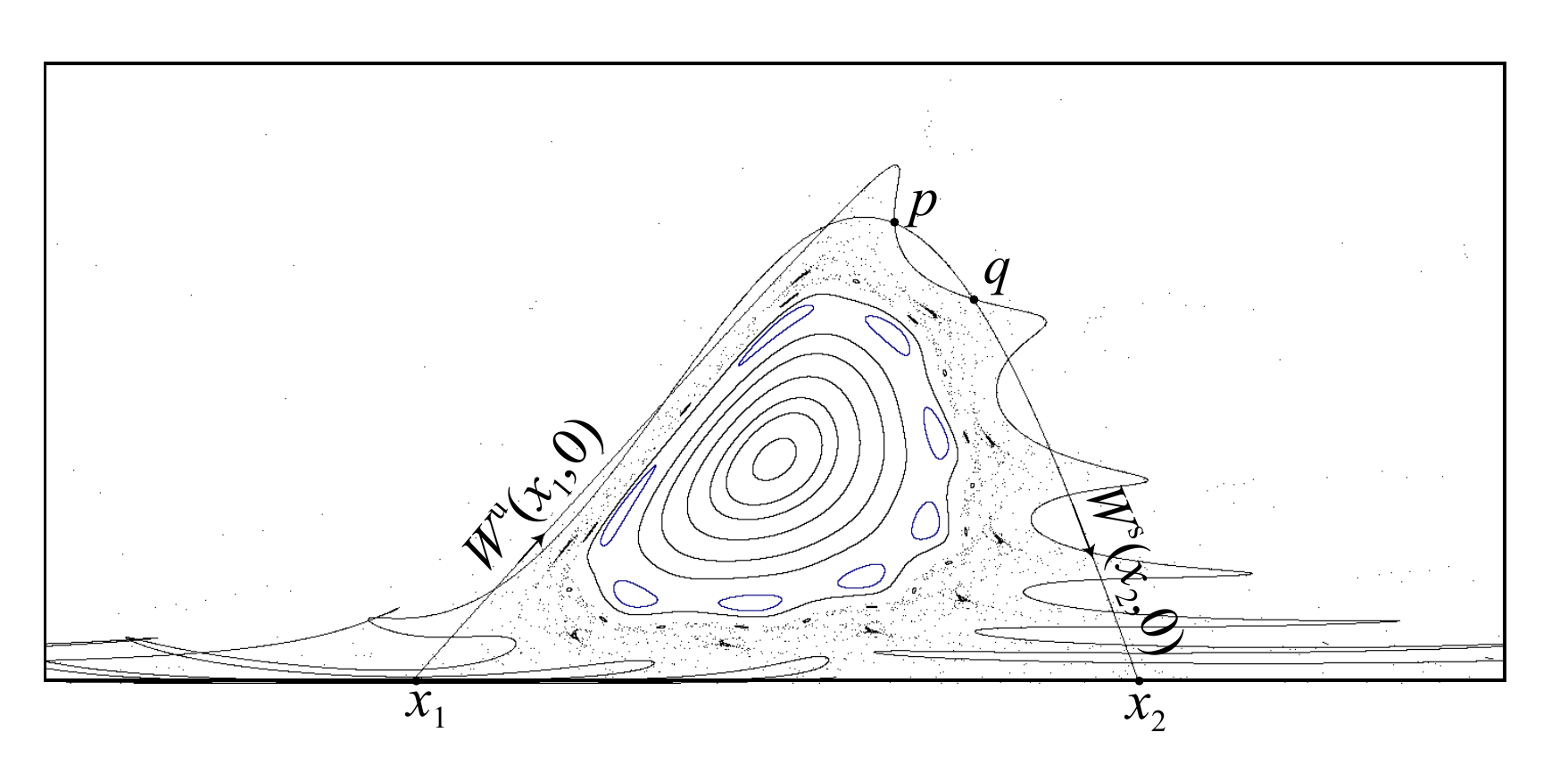}
\caption{Phase space of the map \Eq{eq:HermanMap} for $h(x) = x -\frac{a}{2\pi}
\cos(2\pi x)$ with $a = 0.85$. Here the saddle points lie at $x_1 = -x_2 = -0.25$
and there is an elliptic fixed point at $(0,\frac{a}{2\pi})$.}
\label{fig:hermanMap}
\end{center}
\end{figure}

Note that the line $y = 0$ is invariant under $g$, and the dynamics on this line
is simply $x \mapsto h(x)$. Thus if $h$ has exactly two fixed points, say $x_1 < x_2$,
then $g$ has saddle fixed points at $(x_i, 0)$. Two of the manifolds of these
saddles simply lie on the $x$-axis; for example, if $0 < h'(x_1)  < 1 < h'(x_2)$
then $\Ws(x_1,0) = (-\infty,x_2)$ and $\Wu(x_2,0) = (x_1,\infty)$. The other two
manifolds, $\Wu(x_1,0)$ and $\Ws(x_2,0)$, may also intersect
\cite{tabacman95}. and give rise to a nontrivial resonance zone, see for example
\Fig{fig:hermanMap}. For this example, there are exactly two primary intersection
orbits in the upper-half plane, labelled $p$ and $q$ in the figure.

The three-dimensional map $f$ defined through \Eq{eq:semiD} has an invariant plane
$\left\{y=0\right\}$ and, when $h$ has two fixed points, has a pair of hyperbolic
invariant circles
\[
    \cC_i =\left\{ (x_{i},0,\theta ): \theta \in \Sset\right\} \;, \quad i = 1,2 \;.
\]
The set $\Sigma = \{(x,0,\theta): x_1 < x < x_2 \} = \Ws(\cC_1) \cap \Wu(\cC_2)$
is a heteroclinic manifold that has zero flux. The manifolds $\Wu(\cC_1)$ and
$\Ws(\cC_2)$ can be used to form the second, leaky boundary of a resonance zone.
Indeed, if $p \in \Wu(\cC_1) \cap\Ws(\cC_2)$ is a primary intersection point for
$g$, recall \Fig{fig:hermanMap}, then the circle $\eta_p = \{p\} \times \Sset$ is
a proper boundary for both $\Wu(\cC_1)$ and $\Ws(\cC_2)$. We can then form a
resonance zone $\cR$ with the boundary
\[
    \partial \cR = \Sigma \cup \Wu_{f(\eta_p)}(\cC_1) \cup \Ws_{f(\eta_p)}(\cC_2)
\]
Notice that in this case, no ``caps'' are needed.
If, as in \Fig{fig:hermanMap}, $g$ has exactly one more primary intersection
point, $q$, the set of primary heteroclinic intersections between $\cC_1$ and
$\cC_2$ is given by \Eq{eq:primaryPQ}. Finally, by \Eq{eq:semiDVolume} and 
\Eq{eq:paupau}, the volume of the exit lobe is
\[
    \volume{\bar\cE} =\pi \sum_{k\in \mathbb{Z}}\left((y^q_k)^2-( y^p_k)^2\right) \;.
\]

\subsection{Nonautonomous Hamiltonian flow}\label{sec:Hamiltonian}

The flow of an incompressible vector field is volume-preserving. If the vector
field is exact, in the sense we state below, then its flow will be exact as well.
In this section we compute the volume of lobes that are obtained from the time-$T$
map of such a flow. For this case, our results reduce to those of \cite{MacKay94}.

Recall that an incompressible vector field  $X$ satisfies $L_X\volform{M} \equiv
(\nabla \cdot X) \volform{M} = 0$, where $L_X$ is the Lie derivative. By
\Eq{eq:LieIdentity}, $X$ is incompressible if the form $i_X \volform{M}$ is closed.
Consequently, it is natural to say that $X$ is exact-incompressible when
$i_X\volform{M}$ is exact. Indeed this implies that the flow of the vector field is
exact volume-preserving.

\begin{pro}[\cite{Lomeli08b}]\label{pro:flow}
Suppose $X: M \to TM$ is a vector field with complete flow $\pht$ and  $\volform{M} =
d\alpha$ is an exact volume form on $M$. If $i_X\volform{M}$ is exact, then $L_X \alpha
= d \beta$ for some $n-2$ form $\beta$,  
and the flow is exact volume preserving, $\pht^*\alpha-\alpha=
d\lambda_t$ with the $(n-2)$-form
\begin{equation}\label{eq:lambdaT}
    \lambda_t=\int_0^t \varphi_\tau^*\beta\,d\tau \;,
\end{equation}
for each $t\in \Rset$.
\end{pro}

As an example, consider the case of a $1+\frac12$ degree-of-freedom Hamiltonian flow,
generated by a $C^2$ function $H(x,y,t)$. On the extended phase space $z =
(x,y,\theta) \in \Rset^3$ $H$ generates the vector field
\begin{equation}\label{eq:sistema}
   X = \left(\frac{\partial H}{\partial y},
               -\frac{\partial H}{\partial x}, 1 \right) \;.
\end{equation}
For the volume form $\volform{M}= dx\wedge dy \wedge d\theta$ we can choose $\alpha= -y
dx \wedge d\theta $ so that $d\alpha=\volform{M}$. Hence,
\begin{align*}
    i_X \volform{M} &= dH \wedge d\theta + dx \wedge dy \;, \\
    i_X \alpha &= -y\frac{\partial H}{\partial y}d\theta + y dx \;.
\end{align*}
These imply
\[
    L_X\alpha=i_X\volform{M} + di_X\alpha
             =d\left(H-y\frac{\partial H}{\partial y} \right)\wedge d\theta \;,
\]
so that we can define $\beta= - \cL d\theta $ where
\[
    \cL=y \partial_y H - H
\]
is the (phase space) Lagrangian. Assuming that the flow $\varphi_t$ of
the vector field $X$ is complete, then \Pro{pro:flow} implies that it is exact volume
preserving with
\begin{equation}\label{eq:lagrangian}
    \lambda_t= - \left(\int_0^t \cL\circ\varphi_\tau\,d\tau\right) d\theta \;,
\end{equation}
by \Eq{eq:lambdaT}. Thus $\lambda_t(X)$ is the (negative of the) \emph{action} of
the orbit segment from $\tau = 0$ to $t$.

Now suppose that $H$ is $1-$periodic in its last coordinate:
$H(x,y,\theta+1)=H(x,y,\theta)$, and let $M = \Rset^2 \times \Tset$,
where
$\Tset=\Rset/\Zset$. 

\begin{figure}[th]
\begin{center}
\includegraphics[height=4in]{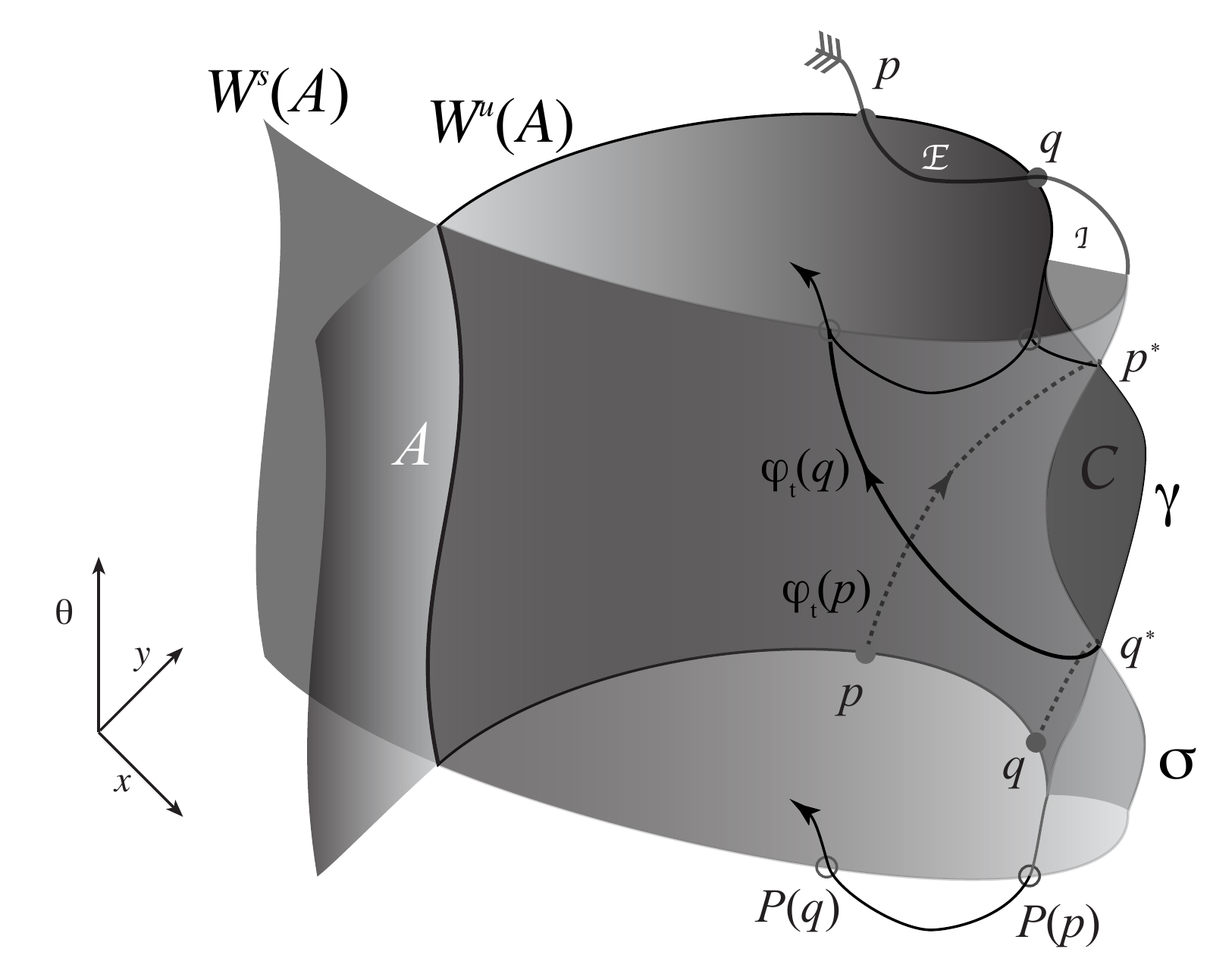}
\caption{Resonance zone for a periodically time dependent flow with a hyperbolic
invariant circle $A$ and two primary intersection orbits $\chi^\pm$.}
\label{fig:3DFlow}
\end{center}
\end{figure}

The Poincar\'e return map to the section $\theta = 0$ is
\[
    P(x,y) = \varphi_1(x,y, 0) \;.
\]
We will assume that $P$ has a saddle fixed point, $a$, whose stable and unstable
manifolds intersect and have exactly two transversal, primary intersection orbits,
e.g., the orbits of $p$ and $q$ as shown in \Fig{fig:2DResonanceStd}. This implies
that the vector field \Eq{eq:sistema} has a hyperbolic invariant circle $A =
\{\varphi_t(a,0):  0 \le t < 1\}$ whose two-dimensional stable and unstable
manifolds have two, primary intersection orbits, $\varphi_t(q)$ and $\varphi_t(p)$
as sketched in \Fig{fig:3DFlow}.

By reducing to the Poincar\'e section, the resonance zone and its exit and
incoming sets for this three-dimensional system can be obtained purely by
considering the area-preserving map $P$. In this case, we can compute the area of
the exit lobe using the standard theory of \cite{MMP84}, to obtain
\begin{equation}\label{eq:2Dvol}
    \area{\cE} = \Delta W
\end{equation}
where $\Delta W$ is the difference between the actions of the two primary
homoclinic orbits:
\begin{equation}\label{eq:deltaW}
    \Delta W = \int_{-\infty}^\infty \left(\cL(\varphi_\tau(q)) 
- \cL(\varphi_\tau(p)) \right) d\tau
\end{equation}
Note that since $P$ is the time-one map of $X$,  $\area{\cE}$ is the area
\emph{per unit time} that escapes from the resonance zone.

To make a test case for \Thm{thm:main}, we now consider the three-dimensional map
\[
    f(x,y,\theta) \equiv \varphi_T(x,y,\theta)
\]
for $T \not\in \Zset$. By \Pro{pro:flow}, $f$ is an exact volume-preserving
diffeomorphism of $M$ with one form $\lambda_T$ given by \Eq{eq:lagrangian}. Under
the above assumptions, $A$ is a normally hyperbolic invariant circle of the map
$f$, and $f$ satisfies \Hyp{1}-\Hyp{3}. 

To construct a resonance zone, $\bar \cR$,
we must introduce caps: since the orbits of $p$ and $q$ are homotopic to $A$, they
cross any proper loops $\sigma$ and $\gamma$.  As usual we select a pair of proper
loops $\gamma$ and $\sigma$, their corresponding fundamental domains $\cU$ and
$\cS$, and a cap $\cC$ obeying \Hyp{4}-\Hyp{6}. We let $\bar\cE$ denote the
three-dimensional exit lobe for this system.  Using the form \Eq{eq:lagrangian},
since $\varphi_t^* d\theta = d\theta$ we have
\begin{equation}\label{eq:fstar}
    (f^k)^* \lambda_T = -(\varphi_{kT})^* \left( \int_0^T \cL \circ \varphi_\tau d\tau\right) d\theta
        = -\left(\int_{kT}^{(k+1)T} \cL \circ \varphi_{\tau} d\tau \right) d\theta
\end{equation}
The set $\eta \subset \cP(A,A)$ used in \Thm{thm:main} corresponds to the primary
intersections on the fundamental domains, thus, $\eta = \eta_p-\eta_q$ where
\begin{align*}
    \eta_p &= \{\varphi_t(p): t \in \Rset\}\cap\cU = \{\varphi_s(p^*): 0 \le s \le T \} \;, \\
    \eta_q &= \{\varphi_t(q): t \in \Rset\}\cap\cU = \{\varphi_s(q^*): 0 \le s \le T \}\;,
\end{align*}
and $\{p^*,q^*\} = \gamma \cap \sigma$ are points on the orbits of $p$ and $q$.

Theorem \ref{thm:main} implies that, in order to compute the volume,
 we must integrate the one form given in \Eq{eq:fstar} over $\eta$ and sum over
$k\in\Zset$. Using the temporal parametrization for $\eta_p$ gives
$d\theta|_{\eta_p} = ds$, and
\[
    \int_{\eta_p}(f^k)^*\lambda_T
      = -\int_0^T \left( \int_{kT}^{(k+1)T} \cL(\varphi_{t+s}(p^*)) dt
          \right) ds
      = -T  \int_{kT}^{(k+1)T} \cL(\varphi_\tau(p^*))d\tau \;,
\]
where we defined a new integration variable $\tau = t+s$ and used the periodicity
of $\cL$ in time. The volume of the exit lobe for $f$ is given by
\Eq{eq:lobeVolume}, and the sum becomes a single integral:
\[
    \volume{\bar\cE} = -T \int_{-\infty}^\infty
      \left(\cL(\varphi_t(p^*))-\cL(\varphi_t(q^*))\right) dt \;.
\]
This integral is geometrically convergent since the orbits of $p^*$ and $q^*$ are
bi-asymptotic to the hyperbolic circle $A$. Moreover, we can replace the initial
conditions with $p$, and $q$, and thus obtain
\begin{equation}\label{eq:3Dvol}
    \volume{\bar\cE} =  T \Delta W \;,
\end{equation}
where $\Delta W$ is given by \Eq{eq:deltaW}. Note that $\volume{\bar\cE}$ is the
volume that exits from the three-dimensional resonance per step of the map $f$,
that is per $T$ units of time. Thus, \Eq{eq:3Dvol} is exactly what is expected
from \Eq{eq:2Dvol}.

\section{Conclusion and future research}

The computation of lobe volumes is a first step toward developing a theory of
transport for volume-preserving maps. These maps appear to have all sticky regions surrounding invariant tori and algebraic decay of exit and transit time distributions, complications that are familiar from the study of area-preserving maps \cite{Meiss92}. Since volume-preserving dynamics pertains to the motion of Lagrangian tracers in incompressible fluids, a theory of transport should prove useful for the understanding laminar mixing and for designing an
optimal mixer \cite{Balasuriya05}.

To apply our result \Eq{eq:lobeVolume} to a general map we must compute the
primary intersection $\cP(A,B)$ between a pair of invariant sets $A$ and $B$.
Finding this set is easiest when the map $f$ is reversible, $R \circ f = f^{-1}
\circ R$, and when $R(A) = B$ for the reversor $R$. In this case, it is easy to
see that if the manifold $\Ws(A)$ intersects the fixed set of the reversor,
$\Fix{R}$, the intersection point must be a point in the primary intersection set
$\cP$. Thus this point can be used as a starting point for a continuation method
to obtain $\eta$. The intersection with $\Fix{f\circ R}$ will give a point on a
second component of $\eta$. An suitable example to study is the quadratic map of
\cite{Lomeli98a}, which is reversible in special cases. We hope to report such
computations in a future paper.


\appendix

\section{Some Notation}\label{app:appendix}
Here we set out our notation, which follows e.g. \cite{Abraham78}. If $\alpha$ is
a $k$-form and $V_1, V_2, \ldots V_k$ are vector fields, then the pullback, $f^*$,
of a diffeomorphism $f$ is defined by
\begin{equation}\label{eq:pullbackForm}
    (f^*\alpha)_x(V_1,V_2,...,V_k) =\alpha_{f(x)}(Df(x)V_1(x),\ldots, Df(x)V_k(x)) \;.
\end{equation}
The pullback can be applied to a vector field $V$ as well:
\begin{equation}\label{eq:pullbackV}
    (f^{*}V)(x) = (Df(x))^{-1}V(f(x)) \;.
\end{equation}
The pushforward operator is defined as
\[
    f_* = (f^{-1})^* \;.
\]
The inner product of $\alpha$ with $V$ is defined as the $(k-1)$-form
\begin{equation}\label{eq:innerProduct}
    i_V \alpha = \alpha(V,\cdot,\ldots,\cdot) \;.
\end{equation}
Suppose that $\varphi_t$ is the ($C^1$) flow of a vector field $V$, so that
$\varphi_0(x) = x$, and $d/dt \varphi_t(x) = V(\varphi_t(x))$. Then the Lie
derivative with respect to $V$ is the differential operator defined by
\begin{equation}\label{eq:LieDeriv}
    L_V \cdot \equiv  \left. \frac{d}{dt}\right|_{t=0}  \varphi_t^* \cdot
\end{equation}
where $\cdot$ is any tensor. The key identity for the derivative is Cartan's
magic formula:
\begin{equation}\label{eq:LieIdentity}
    L_V  \equiv i_V ( d ) + d(i_V ) 
\end{equation}



\end{document}